\documentclass[10pt,oneside,final]{IEEEtran}

\usepackage{amsmath}
\usepackage{amssymb}
\usepackage{bm}
\usepackage{graphicx}
\usepackage{cite}
\usepackage{mathtools}
\usepackage{algorithm}
\usepackage{algpseudocode}
\usepackage{xcolor}
\usepackage[colorlinks,
linkcolor=blue,
urlcolor=black,
anchorcolor=blue,
citecolor=blue]{hyperref}

\AtBeginDocument{%
  \setlength{\abovedisplayskip}{4.8pt plus 2pt minus 1pt}%
  \setlength{\belowdisplayskip}{4.8pt plus 2pt minus 1pt}%
  \setlength{\abovedisplayshortskip}{0pt plus 1pt}%
  \setlength{\belowdisplayshortskip}{2.5pt plus 1pt minus 1pt}%
}

\DeclareMathOperator{\Tr}{Tr}
\DeclareMathOperator{\diag}{diag}
\DeclareMathOperator{\atanTwo}{atan2}
\DeclareMathOperator{\modulo}{mod}

\begin{document}

\title{Practical Hardware-Efficient ISAC: Enabling Multi-Target Sensing With Few RF Chains}

\author{Zhenkun Zhang, Cunhua Pan, Hong Ren, Jun Tang, Dongming Wang, and Jiangzhou Wang,~\IEEEmembership{Fellow,~IEEE}
%\thanks{This work was supported by ...}
\thanks{Zhenkun Zhang, Cunhua Pan, Hong Ren, Jun Tang, Dongming Wang, and Jiangzhou Wang are with National Mobile Communications Research Laboratory, Southeast University, Nanjing 210096, China (e-mail: \{zhenkun\_zhang, cpan, hren, 230268160, wangdm, j.z.wang\}@seu.edu.cn).}
%\thanks{Corresponding authors: Cunhua Pan and Hong Ren.}
}

\maketitle

\newtheorem{remark}{Remark}

\begin{abstract}
Hybrid analog--digital arrays reduce hardware cost in integrated sensing and communications (ISAC), but spatial compression and beam-dependent scan responses complicate multi-target estimation with few radio-frequency (RF) chains. 
To address these challenges, this paper proposes a burst beam-scanning multiple-input multiple-output orthogonal frequency-division multiplexing (MIMO-OFDM) framework that reuses downlink reference signals for joint range, radial-velocity, elevation, and azimuth estimation. 
Specifically, a structured global--local dual-tensor estimator separates targets by using scanning observations and refines their parameters from selected-beam RF-chain responses, with delay--Doppler signatures maintaining cross-stage target correspondence. 
The recovered subarray responses yield closed-form elevation and azimuth estimates through exact sum--difference relations, avoiding small-angle approximations and fine angular grid searches. 
To further improve Doppler accuracy, spatial-mask-aided weighted least squares exploits the longer inter-burst temporal aperture while resolving phase ambiguity using intra-burst estimates. The joint multi-target Cram\'{e}r--Rao lower bound (CRLB) and corresponding three-dimensional (3D) positioning bound are also derived to characterize estimation performance. Simulation results show that the proposed scheme achieves near-CRLB estimation accuracy at high transmit powers and centimeter- to millimeter-level 3D positioning with only four RF chains, while maintaining low computational cost.
\end{abstract}

\begin{IEEEkeywords}
Integrated sensing and communications (ISAC), few-RF-chain base stations, beam scanning, tensor decomposition, multi-target sensing, Cram\'{e}r--Rao lower bound (CRLB).
\end{IEEEkeywords}

%====================================================================
\section{Introduction}

Integrated sensing and communications (ISAC) has been envisioned as a key enabling technology for sixth-generation (6G) wireless networks \cite{liu2022isac,liu2026cooperative}. By sharing spectral resources, transmitted signals, and hardware, ISAC enables cellular infrastructure to provide device-free environmental perception. Low-altitude sensing is an important application that requires accurate three-dimensional (3D) target localization and motion monitoring \cite{tang2025cooperative}. At an individual base station (BS), target range, elevation, and azimuth determine the 3D position, while Doppler measurements provide radial-velocity information. Accurate recovery of these parameters is therefore essential to the sensing capability of cellular BSs.

Orthogonal frequency-division multiplexing (OFDM) \cite{5281762,6094142} supports sensing through the reuse of cellular transmissions. After removing the known transmitted symbols, target delays and Doppler shifts can be derived from the phase progressions across subcarriers and OFDM symbols, respectively \cite{dai2026tutorial}. Our previous work investigated cooperative localization using 5G New Radio (NR) OFDM signals \cite{zhang2025nr} and 3D position and velocity estimation in cooperative bistatic systems \cite{zhang2026bistatic}. These studies established the potential of communication-signal reuse and network cooperation, both of which rely on accurate target measurements supplied by individual sensing receivers. The accuracy of these measurements, in turn, depends on the spatial information retained within the BS receiver architecture.

Hybrid analog--digital arrays support large antenna apertures with a limited number of radio-frequency (RF) chains, reducing hardware cost and power consumption \cite{heath2016mmwave}. However, analog combining compresses the antenna-domain observations and restricts the spatial information available for multi-target estimation. 
To compensate for the limited instantaneous spatial observations of hybrid arrays, beam scanning provides additional measurements by sequentially applying different directional beamforming states. 
Zhang \emph{et al.} developed a multibeam framework that combines fixed communication subbeams with scanning sensing subbeams \cite{zhang2019multibeam}. Resolution-aware scanning further incorporates sensing-resolution and detection requirements into beam design \cite{xu2025resolution}. Recently, Awad \emph{et al.} exploited the beam-sweeping structure of NR synchronization signal bursts for multi-target detection and joint velocity, angle, and range estimation \cite{awad2025ssb}. These studies demonstrated the sensing potential of scanned communication beams. Nevertheless, the temporal evolution of a scanned target response is modulated by direction-dependent beam gains rather than determined by Doppler alone.

Joint processing of frequency, time, and spatial observations enables parameter recovery from hybrid-array measurements. Gupta \emph{et al.} proposed a 3D sparse Bayesian learning method for joint delay, Doppler, and angular estimation in hybrid multiple-input multiple-output (MIMO)-OFDM ISAC systems \cite{gupta2025bayesian}. Its discretized representation links parameter resolution to dictionary size, yet finer grids and extension to both elevation and azimuth increase the computational burden.
This motivates estimation that exploits the available signal structure without a dense joint parameter grid.

Tensor decomposition exploits multidimensional signal structure for joint parameter estimation. Zhang \emph{et al.} developed a unified tensor approach for channel and target estimation and addressed the coupling between departure angle and Doppler induced by the training beams \cite{zhang2024unified}. 
In \cite{zhang2026multibeam}, structured multi-beam tensors and inter-beam correlations are employed to estimate angle and Doppler shift while resolving velocity ambiguity.
These methods demonstrate that beam-dependent angular and Doppler information can be recovered through structured processing. However, their receive-antenna and linear-array formulations cannot directly provide elevation and azimuth recovery from the RF-chain outputs of a subarray-based uniform planar array (UPA). 
Moreover, in OFDM scanning, the scan-domain response contains both direction-dependent gains and Doppler phases, and thus cannot be expressed as a conventional angular array manifold.

For such subarray receivers, target separation and parameter refinement require different information from the scanning observations. 
By jointly processing the scanning observations, the delay--Doppler (DD) signatures of targets can be utilized for multi-target separation, whereas angle recovery relies on the relative phases  between  RF-chains. 
Spatial combining simplifies global processing but discards the inter-chain phase relations required for angle estimation. 
%Target correspondence must therefore be maintained between separated global components and local RF-chain observations. 
Closely spaced intra-burst symbols provide a wide unambiguous Doppler interval but a limited temporal aperture. Inter-burst observations extend this aperture, although their larger sampling interval introduces phase ambiguity and low-gain responses reduce phase reliability. 
Hence, accurate estimation requires consistent use of the global, local, and two-timescale observations.

In this paper, we propose a burst beam-scanning MIMO-OFDM ISAC framework for multi-target sensing at BSs with few RF chains. Downlink OFDM reference signals are transmitted over consecutive symbols under each fixed beamforming state and reused for sensing. A global--local dual-tensor scheme separates targets using scanning observations and refines their parameters using selected-beam RF-chain observations. The main contributions are summarized below.
\begin{itemize}
 \item We establish a burst-scanning observation model for a subarray-based hybrid UPA and develop a global--local dual-tensor estimator. Mode-reduced CANDECOMP/PARAFAC alternating least squares (CP-ALS) exploits the structured delay and intra-burst Doppler factors to extract target-specific DD signatures. A DD-conditioned reprojection (DDRP) scheme is designed to further reconstruct the full-scan target responses from the original observations for beam selection. 
 The global components and locally refined responses are matched by a DD-based association.
 \item Exact subarray sum--difference relations are presented under the adopted array model and obtain closed-form unbiased estimates for the azimuth and elevation.
 The estimator avoids small-angle approximations and fine angular grid searches, enabling continuous dual-angle recovery from a coarse scanning codebook.
 \item A spatial-mask-aided weighted least squares (WLS) Doppler refinement scheme is developed using the reconstructed target-specific inter-burst responses.
 This combines the wider unambiguous Doppler interval of intra-burst sampling with the accuracy gain from inter-burst observations.
 \item We provide the Cram\'{e}r--Rao lower bound (CRLB) analysis for 
 joint multi-target sensing-parameter estimates, accounting for unknown reflection coefficients and cross-target coupling, and obtain the corresponding 3D positioning bound.
 Numerical results demonstrate that the proposed scheme achieves near-CRLB estimation accuracy at high transmit powers and centimeter- to millimeter-level 3D positioning with four RF chains, while maintaining low computational cost.
\end{itemize}

The remainder of this paper is organized as follows. Section~\ref{sec:signal_model} presents the system and signal model. Sections~\ref{sec:global_processing} and \ref{sec:local_refinement} develop global tensor processing and local parameter refinement, respectively. Section~\ref{sec:association_refinement} presents cross-stage association, Doppler refinement, and complexity analysis. Section~\ref{sec:crlb} derives the multi-target CRLB. Numerical results and conclusions are provided in Sections~\ref{sec:simulation} and \ref{sec:conclusion}, respectively.

%====================================================================
\section{System and Signal Model}
\label{sec:signal_model}

\begin{figure}[t]
	\centering
	\includegraphics[width=0.9\linewidth]{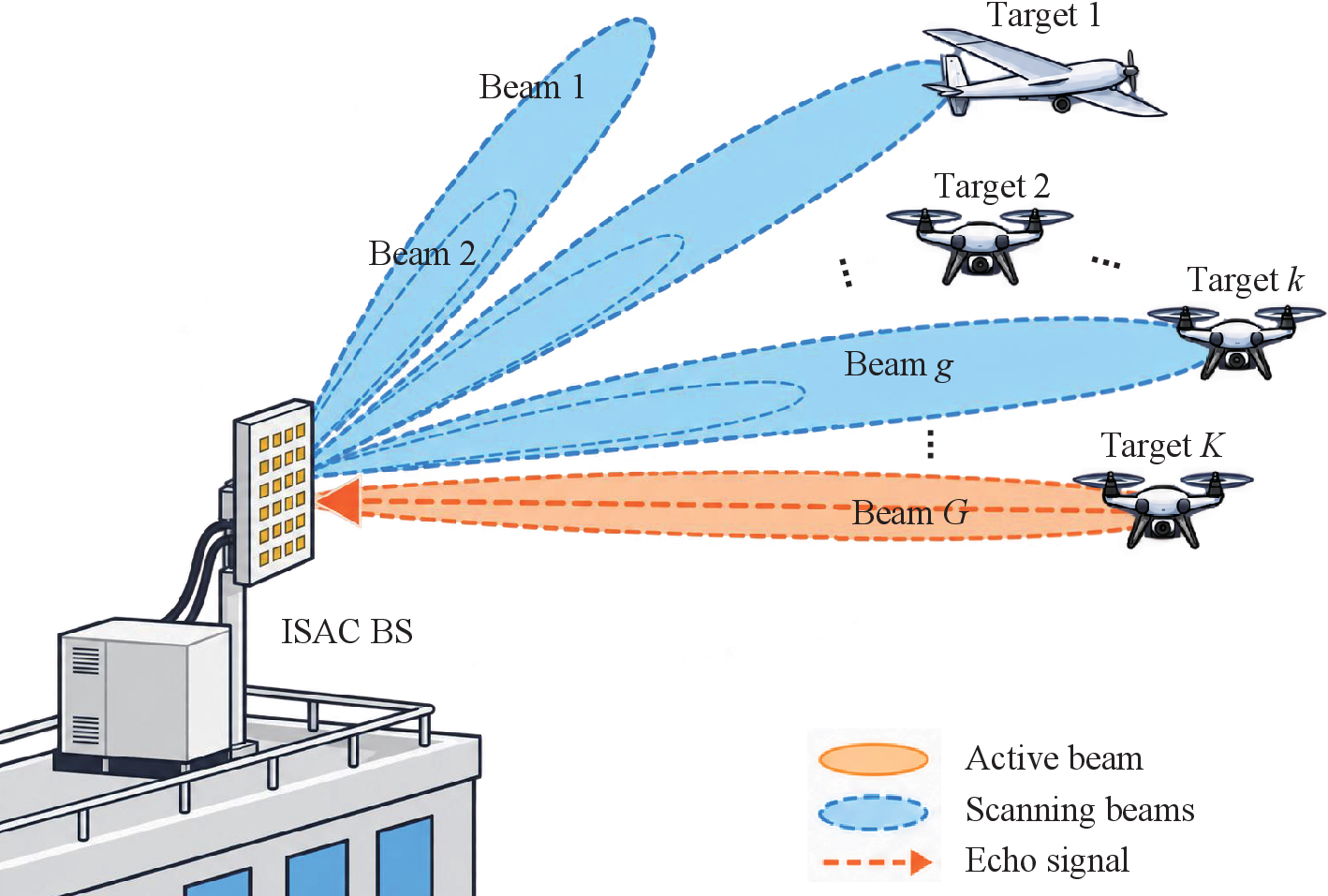}
	\caption{Considered MIMO-OFDM ISAC system that reuses downlink beam-scanning signals for multi-target sensing.}
	\label{fig:systemmodel}
\end{figure}

Consider a low-altitude ISAC system as illustrated in Fig.~\ref{fig:systemmodel}.
The MIMO ISAC BS periodically transmits downlink beam-scanning OFDM signals that serve as channel-estimation reference signals for communication access and are simultaneously reused for environmental sensing.
The sensing objective is to estimate the range, radial velocity, elevation, and azimuth of multiple targets from the echoes collected over one scanning cycle.

To support a large aperture without the cost and power consumption of a fully digital array, the BS employs a subarray-based hybrid analog--digital architecture in which a massive antenna array is connected to only a few radio-frequency (RF) chains. 
Although such architectures are attractive for practical cellular deployments, the limited number of receive RF chains restricts the instantaneous spatial observations available for conventional multi-target sensing. The proposed scheme addresses this limitation by exploiting the beam-scanning process as an additional observation dimension and jointly processing the frequency, beam, OFDM-symbol, and RF-chain dimensions.

\subsection{Subarray Architecture With Few RF Chains }

The BS is equipped with a UPA partitioned into nonoverlapping subarrays\footnote{For analytical tractability, a monostatic architecture is adopted, where one array operates in an in-band full-duplex mode.
	In practical ISAC deployments, a pseudo-monostatic architecture with two identical and closely spaced arrays can be used to alleviate self-interference, with one array used for transmission and the other for reception.
	For far-field targets, the array separation is negligible relative to the target range, and the transmit and receive steering vectors can be approximated by the same array manifold.
	%Accordingly, the pseudo-monostatic configuration is covered by the adopted model after replacing the common steering vector by the corresponding transmit and receive steering vectors when necessary.
}.
Let $2R^{\mathrm{h}}$ and $2R^{\mathrm{v}}$ denote the numbers of subarrays along the horizontal and vertical directions, respectively, where $R^{\mathrm{h}}$ and $R^{\mathrm{v}}$ are positive integers.
The total number of RF chains and subarrays is
\begin{equation}
	R = \left(2R^{\mathrm{h}}\right)\left(2R^{\mathrm{v}}\right).
\end{equation}
Each subarray contains $L^{\mathrm{sub,h}}$ horizontal elements and $L^{\mathrm{sub,v}}$ vertical elements, and is connected to one RF chain.
Defining $L^{\mathrm{sub}}=L^{\mathrm{sub,h}}L^{\mathrm{sub,v}}$, the total number of antenna elements is
\begin{equation}
	L
	=
	\left(2R^{\mathrm{h}}L^{\mathrm{sub,h}}\right)
	\left(2R^{\mathrm{v}}L^{\mathrm{sub,v}}\right)
	=
	R L^{\mathrm{sub}}.
\end{equation}
The inter-element spacings along the horizontal and vertical directions are denoted by $d_{\mathrm{h}}$ and $d_{\mathrm{v}}$, respectively.
Let $f_{\mathrm{c}}$ and $c_0$ denote the carrier frequency and the speed of light, respectively. The carrier wavelength is $\lambda=c_0/f_{\mathrm{c}}$, and the corresponding wavenumber is $\kappa=2\pi/\lambda$.

\begin{figure}[t]
	\centering
	\includegraphics[width=0.7\linewidth]{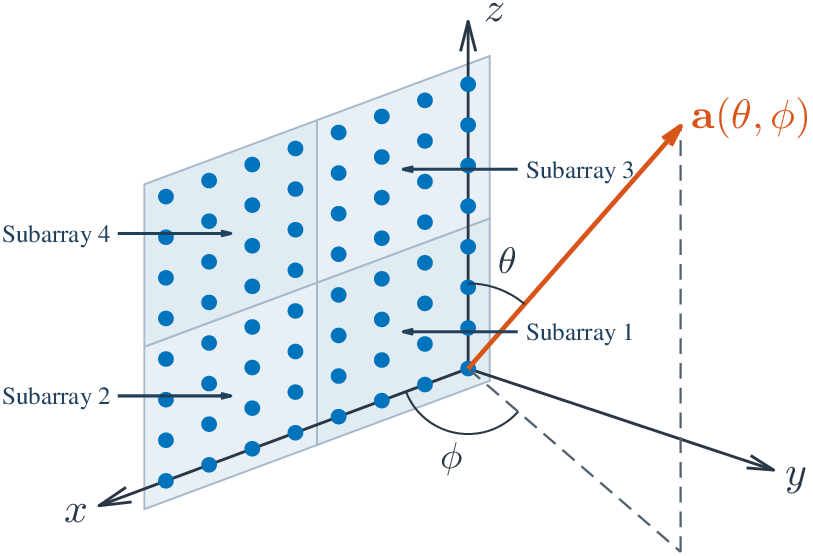}
	\caption{Coordinate system and subarray-based UPA geometry for the minimum-RF-chain configuration with $R=4$.}
	\label{fig:upastructure}
\end{figure}

As shown in Fig.~\ref{fig:upastructure}, the UPA lies in the $x$--$z$ plane and its broadside direction is aligned with the positive $y$-axis.
The origin is placed at the center of the first antenna element.
The elevation angle $\theta\in[0,\pi]$ is measured from the positive $z$-axis, and the azimuth angle $\phi\in[0,\pi)$ is measured from the positive $x$-axis to the projection of the target direction onto the $x$--$y$ plane.

For a target at direction $(\theta,\phi)$, the full-array steering vector is given by\footnote{The adopted horizontal-first vectorization is used throughout the paper.}
\begin{equation}
	\mathbf{a}(\theta,\phi)
	=
	\mathbf{a}^{\mathrm{v}}(\theta)\otimes\mathbf{a}^{\mathrm{h}}(\theta,\phi)
	\in\mathbb{C}^{L\times 1},
\end{equation}
where $\otimes$ denotes the Kronecker product.
The vertical and horizontal steering vectors are given by
\begin{align}
	\mathbf{a}^{\mathrm{v}}(\theta)
	&=
	\left[
	1,
	\ldots,
	e^{j\kappa d_{\mathrm{v}}(2R^{\mathrm{v}}L^{\mathrm{sub,v}}-1)\cos\theta}
	\right]^{\mathrm{T}},
	\\
	\mathbf{a}^{\mathrm{h}}(\theta,\phi)
	&=
	\left[
	1,
	\ldots,
	e^{j\kappa d_{\mathrm{h}}(2R^{\mathrm{h}}L^{\mathrm{sub,h}}-1)\sin\theta\cos\phi}
	\right]^{\mathrm{T}}.
\end{align}

\subsection{Scanning-Beam Design}

Let $\{(\theta_g^{\mathrm{scan}},\phi_g^{\mathrm{scan}})\}_{g=1}^{G}$ denote a predefined 2D scanning grid that covers the angular region of interest,
where the index $g$ represents the temporal execution order of the beam states.
For beam $g$, let
$\boldsymbol{\pi}_g=[p_g^{(1)},p_g^{(2)}]^{\mathrm{T}}\in\mathbb{Z}^{2}$
denote the integer row--column coordinate of $(\theta_g^{\mathrm{scan}},\phi_g^{\mathrm{scan}})$ on the 2D scanning grid.
For the $g$-th scanning direction, the hybrid beamforming matrix is
\begin{equation}
	\mathbf{W}_{g}^{\Sigma}
	=
	\mathbf{W}_{g}^{\Sigma,\mathrm{A}}
	\mathbf{W}_{g}^{\Sigma,\mathrm{D}}
	\in\mathbb{C}^{L\times R}.
\end{equation}
The digital beamformer of all subarrays is selected as
\begin{equation}
	\mathbf{W}_{g}^{\Sigma,\mathrm{D}}
	=
	\frac{1}{\sqrt{L}}\mathbf{I}_{R}.
\end{equation}
Let $\mathcal{I}_r$ denote the set of antenna indices connected to the $r$-th RF chain.
The nonzero entries of the analog beamformer are obtained from the full-array steering vector as
\begin{equation}
	\left[\mathbf{W}_{g}^{\Sigma,\mathrm{A}}\right]_{l,r}
	=
	\left[\mathbf{a}(\theta_g^{\mathrm{scan}},\phi_g^{\mathrm{scan}})\right]_{l},
	\quad
	l\in\mathcal{I}_r,
\end{equation}
where entries outside $\mathcal{I}_r$ are zero.
Since the subarrays are disjoint and contain the same number of elements, we have $\|\mathbf{W}_{g}^{\Sigma}\mathbf{1}_{R}\|_2^2=1$.

\subsection{Sensing Channel and Received-Signal Model}

A burst beam-scanning protocol is adopted in which each burst consists of $Q$ consecutive OFDM symbols transmitted under a fixed beamforming state. The downlink OFDM reference signals are reused for sensing, providing multiple observations for each scanning beam. A guard interval $T_{\mathrm{gap}}$ is reserved between successive bursts for phase-shifter reconfiguration.
Let $T_{\mathrm{s}}$ denote the OFDM-symbol duration. The interval between the starting times of successive bursts is
\begin{equation}
	T_{\mathrm{block}}=Q T_{\mathrm{s}}+T_{\mathrm{gap}}.
\end{equation}
The time stamp of the $q$-th symbol in the $g$-th burst is
\begin{equation}\label{eq:timestamp}
	t_{g,q}
	=
	(g-1)T_{\mathrm{block}}+(q-1)T_{\mathrm{s}},
	\qquad
	q=1,\ldots,Q.
\end{equation}

The BS allocates $M_{\mathrm{c}}$ subcarriers with spacing $\Delta f$ for sensing.
Let $K\ge1$ denote the unknown number of far-field point targets.
We assume that each target's position changes negligibly during the beam scan of a single sensing task, so that its range and angles can be treated as constant over the scan.
The frequency-domain channel matrix on the $m$-th sensing subcarrier is modeled as
\begin{align}\label{H_mgq}
	&\mathbf{H}_{m,g,q}
	=
	\sum_{k=1}^{K}
	\alpha_k
	\mathbf{a}(\theta_k,\phi_k)
	\mathbf{a}^{\mathrm{H}}(\theta_k,\phi_k)
	e^{-j2\pi m\Delta f\tau_k}
	e^{j2\pi f_k^{\mathrm{d}}t_{g,q}},
	\notag \\
	&m=1,\ldots,M_{\mathrm{c}},
\end{align}
where $\alpha_k$, $\tau_k$, and $f_k^{\mathrm{d}}$ denote the two-way complex gain, round-trip delay, and Doppler frequency of target $k$, respectively. With target range $r_k$ and radial velocity $v_k$, the adopted monostatic architecture gives $\tau_k=2r_k/c_0$ and $f_k^{\mathrm{d}}=2v_k/\lambda$.

To facilitate the elimination of reference signals for multi-dimensional parameter extraction, the same unit-modulus symbol $s_{m,g,q}$ is applied to all RF chains, with $|s_{m,g,q}|^2=1$.
The transmitted antenna-domain vector is
\begin{equation}
	\mathbf{x}_{m,g,q}
	=
\sqrt{\frac{P_{\mathrm{t}}}{M_{\mathrm{c}}}}
	\mathbf{W}_{g}^{\Sigma}
	\mathbf{1}_{R}
	s_{m,g,q},
\end{equation}
where $P_{\mathrm{t}}$ is the total transmit power.
After hybrid receive combining, the received vector is
\begin{align}
	\mathbf{y}_{m,g,q}
	&=
	\left(\mathbf{W}_{g}^{\Sigma}\right)^{\mathrm{H}}
	\mathbf{H}_{m,g,q}
	\mathbf{x}_{m,g,q}
	+
	\mathbf{n}_{m,g,q}
	\\
	&=
\sqrt{\frac{P_{\mathrm{t}}}{M_{\mathrm{c}}}}
	\left(\mathbf{W}_{g}^{\Sigma}\right)^{\mathrm{H}}
	\mathbf{H}_{m,g,q}
	\mathbf{W}_{g}^{\Sigma}
	\mathbf{1}_{R}
	s_{m,g,q}
	+
	\mathbf{n}_{m,g,q},\notag
\end{align}
where $\mathbf{n}_{m,g,q}\in\mathbb{C}^{R\times 1}$ denotes zero-mean circularly symmetric complex Gaussian post-combining noise.
Demodulation by the known sensing symbol gives
\begin{align}\label{eq:yhat}
	\hat{\mathbf{y}}_{m,g,q}
	&\triangleq
	\mathbf{y}_{m,g,q}s_{m,g,q}^{*}
	\notag\\
	&=
	\sqrt{\frac{P_{\mathrm{t}}}{M_{\mathrm{c}}}}
	\left(\mathbf{W}_{g}^{\Sigma}\right)^{\mathrm{H}}
	\mathbf{H}_{m,g,q}
	\mathbf{W}_{g}^{\Sigma}
	\mathbf{1}_{R}
	+
	\hat{\mathbf{n}}_{m,g,q},
\end{align}
where $\hat{\mathbf{n}}_{m,g,q}=\mathbf{n}_{m,g,q}s_{m,g,q}^{*}$ has the same distribution as $\mathbf{n}_{m,g,q}$.
For subsequent derivations, we define the RF-chain-domain directional response of target $k$ under beam $g$ as
\begin{equation}\label{b_gk}
	\mathbf{b}_{g,k}
	\triangleq
\sqrt{\frac{P_{\mathrm{t}}}{M_{\mathrm{c}}}}
	\left(\mathbf{W}_{g}^{\Sigma}\right)^{\mathrm{H}}
	\mathbf{a}(\theta_k,\phi_k)
	\mathbf{a}^{\mathrm{H}}(\theta_k,\phi_k)
	\mathbf{W}_{g}^{\Sigma}
	\mathbf{1}_{R}.
\end{equation}
Substituting \eqref{H_mgq} into \eqref{eq:yhat} yields
\begin{equation}\label{eq:yhat_expanded}
	\hat{\mathbf{y}}_{m,g,q}
	=
	\sum_{k=1}^{K}
	\alpha_k
	\mathbf{b}_{g,k}
	e^{-j2\pi m\Delta f\tau_k}
	e^{j2\pi f_k^{\mathrm{d}}t_{g,q}}
	+
	\hat{\mathbf{n}}_{m,g,q}.
\end{equation}

%====================================================================
\section{Global Tensor Processing}
\label{sec:global_processing}

The limited number of RF chains restricts the instantaneous spatial degrees of freedom (DoF) available for multi-target estimation.
The global stage therefore exploits the measurements collected over subcarriers, scanned beams, and OFDM symbols to form a structured low-dimensional tensor representation.
It extracts target-specific DD signatures and recovers their responses over the complete scan grid, thereby providing robust initialization for the local refinement.

\subsection{Spatial Compression and Global-Tensor Construction}

By applying the equal-gain combiner
\begin{equation}
\mathbf{t}
=
\frac{1}{\sqrt{R}}\mathbf{1}_{R},
\end{equation}
we obtain the scalar observation as
\begin{equation}
 z_{m,g,q}
 =
 \mathbf{t}^{\mathrm{H}}\hat{\mathbf{y}}_{m,g,q}.
\end{equation}
Define
\begin{equation}\label{eq:beta_gk}
\beta_{g,k}
\triangleq
\mathbf{t}^{\mathrm{H}}\mathbf{b}_{g,k}
=
\sqrt{\frac{P_{\mathrm{t}}}{RM_{\mathrm{c}}}}
\left|
\mathbf{a}^{\mathrm{H}}(\theta_k,\phi_k)
\mathbf{W}_{g}^{\Sigma}\mathbf{1}_{R}
\right|^2.
\end{equation}
Using \eqref{eq:timestamp}, the scalar observation can be expressed as
\begin{align}\label{eq:z_expanded}
 z_{m,g,q}
 =
 \sum_{k=1}^{K}
 \alpha_k
 &\underbrace{e^{-j2\pi m\Delta f\tau_k}}_{\text{delay factor}}
 \underbrace{\beta_{g,k}e^{j2\pi f_k^{\mathrm{d}}(g-1)T_{\mathrm{block}}}}_{\text{beam-domain factor}}
 \notag\\
 &\times
 \underbrace{e^{j2\pi f_k^{\mathrm{d}}(q-1)T_{\mathrm{s}}}}_{\text{intra-burst Doppler factor}}
 +
 \tilde n_{m,g,q},
\end{align}
where $\tilde n_{m,g,q}=\mathbf{t}^{\mathrm{H}}\hat{\mathbf{n}}_{m,g,q}$.
The burst beam-scanning protocol separates the Doppler phase into inter-burst and intra-burst terms, sampled at intervals $T_{\mathrm{block}}$ and $T_{\mathrm{s}}$, respectively.

Collecting $z_{m,g,q}$ over $(m,g,q)$ yields
$\mathcal{Y}\in\mathbb{C}^{M_{\mathrm{c}}\times G\times Q}$ with the following $K$-component CP model \cite{kolda2009}
\begin{equation}\label{eq:global_tensor}
 \mathcal{Y}
 =
 \sum_{k=1}^{K}
 \alpha_k
 \mathbf{a}_{\tau,k}
 \circ
 \mathbf{a}_{\theta,k}
 \circ
 \mathbf{a}_{\nu,k}
 +
 \mathcal{N},
\end{equation}
where $\circ$ denotes the vector outer product and
\begin{align}
 \mathbf{a}_{\tau,k}
 &=
 \left[
 e^{-j2\pi\Delta f\tau_k},\ldots,
 e^{-j2\pi M_{\mathrm{c}}\Delta f\tau_k}
 \right]^{\mathrm{T}},
 \label{eq_a_tau}
 \\
 \mathbf{a}_{\theta,k}
 &=
 \left[
 \beta_{g,k}e^{j2\pi f_k^{\mathrm{d}}(g-1)T_{\mathrm{block}}}
 \right]_{g=1}^{G},
 \label{eq_a_theta}
 \\
 \mathbf{a}_{\nu,k}
 &=
 \left[
 e^{j2\pi f_k^{\mathrm{d}}(q-1)T_{\mathrm{s}}}
 \right]_{q=1}^{Q}.
 \label{eq_a_nu}
\end{align}
Define the corresponding factor matrices as
\begin{align}
 \mathbf{A}_{\tau}
 &=
 [\mathbf{a}_{\tau,1},\ldots,\mathbf{a}_{\tau,K}]
 \in\mathbb{C}^{M_{\mathrm{c}}\times K},\\
 \mathbf{A}_{\theta}
 &=
 [\mathbf{a}_{\theta,1},\ldots,\mathbf{a}_{\theta,K}]
 \in\mathbb{C}^{G\times K},\\
 \mathbf{A}_{\nu}
 &=
 [\mathbf{a}_{\nu,1},\ldots,\mathbf{a}_{\nu,K}]
 \in\mathbb{C}^{Q\times K}.
\end{align}
The third tensor mode is indexed by the OFDM symbols within each burst and is referred to as the intra-burst symbol dimension. The second mode is indexed by the scanning beams, with one burst corresponding to each beam state.
The delay and intra-burst Doppler factors are Vandermonde. In contrast, the beam-domain factor contains both the inter-burst Doppler phase and the direction-dependent beam gain $\beta_{g,k}$, and is therefore generally non-Vandermonde.

\subsection{Information-Theoretic-Based Target-Number Estimation}

The target number is estimated before CP decomposition using the information-theoretic model-order framework \cite{wax1985}.
The mode-2 unfolding of \eqref{eq:global_tensor} is
\begin{equation}\label{eq:Y2_unfold}
 \mathbf{Y}_{(2)}
 =
 \mathbf{A}_{\theta}
 \diag(\boldsymbol{\alpha})
 (\mathbf{A}_{\nu}\odot\mathbf{A}_{\tau})^{\mathrm{T}}
 +
 \mathbf{N}_{(2)},
\end{equation}
where $\boldsymbol{\alpha}=[\alpha_1,\ldots,\alpha_K]^{\mathrm{T}}$ and $\odot$ denotes the Khatri--Rao product.
Let $N=M_{\mathrm{c}}Q$, $D=\min(G,N)$, and $N_{\mathrm{s}}=\max(G,N)$.
The compact covariance matrix is formed as
\begin{equation}\label{eq:compact_cov}
 \hat{\mathbf{R}}_{\mathrm{c}}
 =
 \begin{cases}
 \dfrac{1}{N}\mathbf{Y}_{(2)}\mathbf{Y}_{(2)}^{\mathrm{H}}, & G\le N,\\[1.2ex]
 \dfrac{1}{G}\mathbf{Y}_{(2)}^{\mathrm{H}}\mathbf{Y}_{(2)}, & G>N.
 \end{cases}
\end{equation}
Let $\zeta_1\ge\cdots\ge\zeta_D\ge0$ denote its eigenvalues, and let $K_{\max}$ denote the prescribed maximum target order supported by the estimator. Define $K_{\mathrm{AIC}}=\min\{K_{\max},D-1\}$. For $k\in\{0,\ldots,K_{\mathrm{AIC}}\}$, the Akaike information criterion (AIC) is
\begin{align}
 \operatorname{AIC}(k)
 =
 &-N_{\mathrm{s}}(D-k)
 \ln\left(
 \frac{
 \left(\prod_{i=k+1}^{D}\zeta_i\right)^{1/(D-k)}
 }{
 \dfrac{1}{D-k}\sum_{i=k+1}^{D}\zeta_i
 }
 \right)
 \notag\\
 &+k(2D-k).
 \label{eq:AIC}
\end{align}
Then, the estimated target number is
\begin{equation}
 \hat K
 =
 \mathrm{arg}\min_{1\le k\le K_{\mathrm{AIC}}} \mathrm{AIC}\left( k \right) .
\end{equation}

\subsection{2D Spatial Beam Screening}

Only a small fraction of scanning beams usually contains appreciable target energy.
Retaining all $G$ beam slices in the global CP decomposition increases both the computational cost and the number of noise-dominated entries in the beam-domain factor.
We therefore screen the beam dimension before decomposition.

Define the raw energy of the $g$-th beam slice as
\begin{equation}
 E_g^{(0)}
 =
 \left\|[\mathcal{Y}]_{:,g,:}\right\|_{\mathrm{F}}^2.
\end{equation}
The energies are placed at the grid locations $\{\boldsymbol{\pi}_g\}$ and spatially pooled using a nonnegative local smoothing kernel, e.g., the $3\times3$ binomial kernel
\begin{equation}
 \mathbf{H}_{\mathrm{sp}}
 =
 \begin{bmatrix}
 0.25&0.5&0.25\\
 0.5&1&0.5\\
 0.25&0.5&0.25
 \end{bmatrix}.
\end{equation}
Let $N_{\mathrm{beam}}$ denote the expected beam-footprint size per target and let $G_{\min}$ denote a prescribed minimum number of retained beams.
The number of retained beams is selected as
\begin{equation}\label{eq:Gkeep}
 G_{\mathrm{keep}}
 =
 \min\left\{
 G,
 \max\left(G_{\min},N_{\mathrm{beam}}\hat K\right)
 \right\}.
\end{equation}
The $G_{\mathrm{keep}}$ beams with the largest pooled energies are retained. The screened tensor is then obtained as
\begin{equation}
 \mathcal{Y}_{\mathrm{keep}}
 =
 \mathcal{Y}(:,\mathcal{G}_{\mathrm{keep}},:)
 \in
 \mathbb{C}^{M_{\mathrm{c}}\times G_{\mathrm{keep}}\times Q},
\end{equation}
where $\mathcal{G}_{\mathrm{keep}}=\{g_1,\ldots,g_{G_{\mathrm{keep}}}\}$
denotes the retained original beam indices, arranged according to their physical scanning times.

\subsection{Noise-Gated Spatial Dynamic-Range Compression}

Multi-target echoes may exhibit a significant near-far effect.
A strong target can produce strong echoes under several neighboring beam states owing to the finite mainlobe width and sidelobes, thereby obscuring a weaker target in the global decomposition.
To reduce this imbalance without amplifying noise-dominated slices, a noise-gated nonlinear compression is applied to $\mathcal{Y}_{\mathrm{keep}}$.

Define
\begin{equation}
 E_g
 =
 \left\|[\mathcal{Y}_{\mathrm{keep}}]_{:,g,:}\right\|_{\mathrm{F}},
 \quad g=1,\ldots,G_{\mathrm{keep}},
\end{equation}
and let $E_{\mathrm{floor}}$ be the mean of the smallest $\lceil G_{\mathrm{keep}}/2\rceil$ values in $\{E_g\}$.
The compression weights are
\begin{equation}\label{eq:compression_weight}
 \!\!w_g
 =
 \begin{cases}
 \left(\dfrac{E_g}{E_{\mathrm{floor}}}\right)^{\gamma}, & \!\!E_g>E_{\mathrm{floor}},\\
 1, & \!\!E_g\le E_{\mathrm{floor}},
 \end{cases}
 \quad
 g=1,\ldots,G_{\mathrm{keep}},
\end{equation}
where $\gamma\in(0,1)$ controls the compression strength.
The resulting tensor is
\begin{equation}\label{eq:Ycomp}
 [\mathcal{Y}_{\mathrm{cmp}}]_{:,g,:}
 =
 \frac{[\mathcal{Y}_{\mathrm{keep}}]_{:,g,:}}{w_g}.
\end{equation}

\subsection{Mode-Reduced Regularized CP-ALS}

Directly computing the CP decomposition of $\mathcal{Y}_{\mathrm{cmp}}$ via ALS is computationally inefficient because the subcarrier mode is typically much larger than the beam and symbol modes.
We therefore firstly compress the mode-1 signal subspace by a truncated higher-order singular value decomposition (HOSVD), which preserves the dominant low-rank structure while reducing the dimensions of the ALS iterations \cite{delathauwer2000,kolda2009}.
Let
$\mathbf{Y}_{\mathrm{cmp},(1)}\in\mathbb{C}^{M_{\mathrm{c}}\times G_{\mathrm{keep}}Q}$
be the mode-1 unfolding of $\mathcal{Y}_{\mathrm{cmp}}$.
With a small rank buffer $p_{\mathrm{buf}}\ge0$, define
\begin{equation}
 R_1
 =
 \min\{M_{\mathrm{c}},G_{\mathrm{keep}}Q,\hat K+p_{\mathrm{buf}}\}.
\end{equation}
The $R_1$ dominant left singular vectors of $\mathbf{Y}_{\mathrm{cmp},(1)}$ form
$\mathbf{U}_1\in\mathbb{C}^{M_{\mathrm{c}}\times R_1}$.
Using $\times_1$ to denote the mode-1 tensor--matrix product, the reduced core tensor is
\begin{equation}\label{eq:Ycore}
 \mathcal{Y}_{\mathrm{core}}
 =
 \mathcal{Y}_{\mathrm{cmp}}\times_1\mathbf{U}_1^{\mathrm{H}}
 \in
 \mathbb{C}^{R_1\times G_{\mathrm{keep}}\times Q}.
\end{equation}
The corresponding factor matrices are
$\mathbf{A}_{\tau,\mathrm{c}}=\mathbf{U}_1^{\mathrm{H}}\mathbf{A}_{\tau}$,
$\mathbf{A}_{\theta,\mathrm{cmp}}\in\mathbb{C}^{G_{\mathrm{keep}}\times K}$,
and $\mathbf{A}_{\nu}$, where $\mathbf{A}_{\theta,\mathrm{cmp}}$ denotes the beam-domain factor after beam screening and the slice-wise scaling in \eqref{eq:Ycomp}.
Thus, the core-tensor CP model is parameterized by the factor matrices
$\{\mathbf{A}_{\tau,\mathrm{c}},\mathbf{A}_{\theta,\mathrm{cmp}},\mathbf{A}_{\nu}\}$.
For CP-ALS, the complex gains are absorbed into the beam-factor columns, and all factor matrices are fitted with $\hat K$ columns.

When targets are closely spaced, the columns of these factor matrices can become highly correlated, leading to ill-conditioned ALS normal equations and unstable updates.
We therefore introduce a scale-adaptive diagonal Tikhonov term into each mode update \cite{tikhonov1977}.
For the delay mode, define
$\mathbf{Z}_{\tau}=\mathbf{A}_{\nu}\odot\mathbf{A}_{\theta,\mathrm{cmp}}$.
The regularized update solves the following problem
\begin{equation}\label{eq:reg_als_obj}
 \min_{\mathbf{A}_{\tau,\mathrm{c}}}
 \quad
 \left\|
 \mathbf{Y}_{\mathrm{core},(1)}
 -
 \mathbf{A}_{\tau,\mathrm{c}}\mathbf{Z}_{\tau}^{\mathrm{T}}
 \right\|_{\mathrm{F}}^2
 +
 \Tr\left(
 \mathbf{A}_{\tau,\mathrm{c}}
 \mathbf{R}_{\tau}
 \mathbf{A}_{\tau,\mathrm{c}}^{\mathrm{H}}
 \right),
\end{equation}
where
\begin{align}
 \mathbf{V}_{\tau}
 &=
 \mathbf{Z}_{\tau}^{\mathrm{H}}\mathbf{Z}_{\tau},\\
 \mathbf{R}_{\tau}
 &=
 \lambda_{\mathrm{Tik}}
 \diag\!\left(\diag(\mathbf{V}_{\tau})\right),
\end{align}
with $\lambda_{\mathrm{Tik}}>0$ denoting a small regularization coefficient.
The resulting update is
\begin{equation}\label{eq:reg_als_update}
 \mathbf{A}_{\tau,\mathrm{c}}
 =
 \mathbf{Y}_{\mathrm{core},(1)}
 \mathbf{Z}_{\tau}^{*}
 \left(
 \mathbf{V}_{\tau}^{\mathrm{T}}+\mathbf{R}_{\tau}
 \right)^{-1}.
\end{equation}
The beam-domain factor $\mathbf{A}_{\theta,\mathrm{cmp}}$ and the intra-burst Doppler factor $\mathbf{A}_{\nu}$ are updated analogously by cyclic permutation of the tensor modes, using the same diagonal regularization form.
After convergence, the delay factor is lifted back to the original subcarrier domain as
\begin{equation}\label{eq:hat_A_tau_raw}
 \hat{\mathbf{A}}_{\tau,\mathrm{raw}}
 =
 \mathbf{U}_1\hat{\mathbf{A}}_{\tau,\mathrm{c}},
\end{equation}
whereas $\hat{\mathbf{A}}_{\nu,\mathrm{raw}}$ is obtained directly from the third mode of the core decomposition.

The physical delay and intra-burst Doppler factors in \eqref{eq_a_tau} and \eqref{eq_a_nu} have unit-modulus Vandermonde structure, which is not enforced by the unconstrained ALS updates.
To restore this physical structure, each recovered delay or Doppler column is projected onto its corresponding Vandermonde manifold.
Consider a generic factor
$\mathbf{x}=[x_1,\ldots,x_{N_{\mathrm{x}}}]^{\mathrm{T}}$
of length $N_{\mathrm{x}}$, associated with the consecutive integer index vector
$\mathbf{n}=[n_0,n_0+1,\ldots,n_0+N_{\mathrm{x}}-1]^{\mathrm{T}}$.
For a nonzero complex input $z$, define the principal-phase operator as
\begin{equation}\label{eq:phase_unwrap}
 \mathcal{U}(z)
 \triangleq
 \atanTwo\!\left(\Im\{z\},\Re\{z\}\right)
 \in(-\pi,\pi].
\end{equation}
Thus, $\mathcal{U}(z)$ extracts the phase of $z$ in the interval $(-\pi,\pi]$.
A robust coarse phase increment is then obtained from adjacent-sample correlation as
\begin{equation}\label{eq:strict_coarse_slope}
 \psi^{(0)}
 =
 \mathcal{U}\!\left(
\sum_{\ell=1}^{N_{\mathrm{x}}-1}x_{\ell}^{*}x_{\ell+1}
 \right).
\end{equation}
After removing this dominant phase progression,
\begin{equation}\label{eq:strict_derotation}
 \tilde x_{\ell}
 =
 x_{\ell}e^{-j\psi^{(0)}n_{\ell}},
\qquad \ell=1,\ldots,N_{\mathrm{x}},
\end{equation}
the residual phase sequence is unwrapped recursively as
\begin{align}
 \varphi_1
 &=
 \mathcal{U}(\tilde x_1),\notag\\
 \varphi_{\ell}
 &=
 \varphi_{\ell-1}
 +
 \mathcal{U}\!\left(\tilde x_{\ell-1}^{*}\tilde x_{\ell}\right),
\quad \ell=2,\ldots,N_{\mathrm{x}}.
 \label{eq:phase_unwrap_recursion}
\end{align}
A Tretter-type linear regression \cite{tretter1985} then yields the residual phase slope
\begin{equation}\label{eq:strict_residual_slope}
 \Delta\psi
 =
 \frac{
\sum_{\ell=1}^{N_{\mathrm{x}}}(n_{\ell}-\bar n)(\varphi_{\ell}-\bar\varphi)
 }{
\sum_{\ell=1}^{N_{\mathrm{x}}}(n_{\ell}-\bar n)^2
 },
\end{equation}
where $n_{\ell}=n_0+\ell-1$, and $\bar n$ and $\bar\varphi$ denote the corresponding sample means.
The refined phase increment is
\begin{equation}
 \hat\psi=\psi^{(0)}+\Delta\psi,
\end{equation}
and the strict Vandermonde reconstruction is defined as
\begin{equation}\label{eq:strict_operator}
 \mathcal{R}_{n_0}(\mathbf{x})
 \triangleq
 \left[
 e^{j\hat\psi n_0},
 \ldots,
e^{j\hat\psi(n_0+N_{\mathrm{x}}-1)}
 \right]^{\mathrm{T}}.
\end{equation}
%The factor amplitude and constant phase offset are omitted in \eqref{eq:strict_operator}, since both can be absorbed by the inherent column-scaling ambiguity of the CP model.
For global component $k$, the strict DD factors are therefore
\begin{align}
 \bar{\mathbf{a}}_{\tau,k}
 &=
 \mathcal{R}_{1}
 \left(\hat{\mathbf{a}}_{\tau,\mathrm{raw},k}\right),
 \label{eq:strict_global_tau}
 \\
 \bar{\mathbf{a}}_{\nu,k}
 &=
 \mathcal{R}_{0}
 \left(\hat{\mathbf{a}}_{\nu,\mathrm{raw},k}\right).
 \label{eq:strict_global_nu}
\end{align}

\subsection{Delay--Doppler-Conditioned All-Beam Reprojection}

Although beam screening and dynamic-range compression improve the robustness of the global decomposition, they discard or rescale part of the beam-domain information. 
To achieve reliable beam selection and inter-burst Doppler refinement, we design a DDRP scheme to recover the target-specific complex responses over the complete scan grid by
conditioning on the strict DD factors in \eqref{eq:strict_global_tau}--\eqref{eq:strict_global_nu} and re-estimating the beam-domain responses from the original all-beam observations.

For target $k$, define the DD signature as
\begin{equation}\label{eq:dd_signature}
 \mathbf{s}_{\mathrm{DD},k}
 =
 \bar{\mathbf{a}}_{\nu,k}\otimes\bar{\mathbf{a}}_{\tau,k}
 \in\mathbb{C}^{M_{\mathrm{c}}Q}.
\end{equation}
Define $\mathbf{Z}_g\in\mathbb{C}^{M_{\mathrm{c}}\times Q}$ by $[\mathbf{Z}_g]_{m,q}=z_{m,g,q}$ and let $\mathbf{z}_g=\operatorname{vec}(\mathbf{Z}_g)$ for beam $g$.
Under the global signal model, we have
\begin{equation}\label{eq:ddrp_model}
 \mathbf{z}_g
 \approx
 \mathbf{S}_{\mathrm{DD}}\mathbf{d}_g
 +
 \mathbf{n}_g,
\end{equation}
where $
\mathbf{S}_{\mathrm{DD}}
=
[\mathbf{s}_{\mathrm{DD},1},\ldots,\mathbf{s}_{\mathrm{DD},\hat K}]
\in\mathbb{C}^{M_{\mathrm{c}}Q\times\hat K}
$
collects the DD signatures,
and the target-specific complex responses $\mathbf{d}_g\in\mathbb{C}^{\hat K}$ under beam $g$ is expressed as 
\begin{equation}\label{eq:d_g}
 \mathbf{d}_g
 =
 \left[
 \alpha_k\beta_{g,k}
 e^{j2\pi f_k^{\mathrm{d}}(g-1)T_{\mathrm{block}}}
 \right]_{k=1}^{\hat K}.
\end{equation}

%Because finite resolution can leave the recovered DD signatures mutually correlated, a ridge-regularized least-squares projection is used to separate their beam-domain responses. 
By stacking $\mathbf{z}_g$ over all beams as $\mathbf{Z}=[\mathbf{z}_1,\ldots,\mathbf{z}_G]\in\mathbb{C}^{M_{\mathrm{c}}Q\times G}$, the full-beam factor is obtained as 
\begin{equation}\label{eq:A_theta_DD}
 \hat{\mathbf{A}}_{\theta}^{\mathrm{DD}}
 =
 \mathbf{Z}^{\mathrm{T}}\mathbf{S}_{\mathrm{DD}}^{*}
 \left(
 \mathbf{S}_{\mathrm{DD}}^{\mathrm{T}}\mathbf{S}_{\mathrm{DD}}^{*}
 +
 \lambda_{\mathrm{DD}}\mathbf{I}_{\hat K}
 \right)^{-1},
\end{equation}
where $\lambda_{\mathrm{DD}}>0$ is a small regularization coefficient. 
Unlike the screened global CP decomposition,  DDRP in \eqref{eq:A_theta_DD} reconstructs one target-specific complex response sequence across all $G$ beams.

For target $k$, define the beam-domain power map as
\begin{equation}
 P_{g,k}^{\mathrm{DD}}
 =
 \left|
 \hat a_{\theta,k}^{\mathrm{DD}}(g)
 \right|^2,
 \qquad g=1,\ldots,G.
\end{equation}
The power map is placed on the 2D scan grid and spatially smoothed using the same local kernel as in the screening stage. Denoting the smoothed values by $\widetilde P_{g,k}^{\mathrm{DD}}$, the optimal beam index associated with target $k$ is selected as
\begin{equation}\label{eq:g_k_DD}
 \hat g_k
 =
 \arg\max_{1\le g\le G}
 \widetilde P_{g,k}^{\mathrm{DD}}.
\end{equation}
%The columns of $ \hat{\mathbf{A}}_{\theta}^{\mathrm{DD}}$, denoted by $\hat{\mathbf{a}}_{\theta,k}^{\mathrm{DD}}$'s, are retained for the fine Doppler refinement described later.

\begin{algorithm}[t]
 \caption{Global Structured DD Extraction and DDRP}
 \label{alg:global}
 \begin{algorithmic}[1]
 \Require Demodulated observations $\hat{\mathbf{y}}_{m,g,q}$ and grid locations $\{\boldsymbol{\pi}_g\}_{g=1}^{G}$.
 \State Obtain $z_{m,g,q}=\mathbf{t}^{\mathrm{H}}\hat{\mathbf{y}}_{m,g,q}$ and construct $\mathcal{Y}$;
 \State Estimate $\hat K$ using \eqref{eq:compact_cov}--\eqref{eq:AIC};
 \State Perform 2D beam screening using \eqref{eq:Gkeep};
 \State Apply noise-gated spatial compression using \eqref{eq:compression_weight}--\eqref{eq:Ycomp};
 \State Construct $\mathcal{Y}_{\mathrm{core}}$ using \eqref{eq:Ycore} and perform regularized CP-ALS;
 \State Recover the raw delay/Doppler factors and strictly rectify them using \eqref{eq:strict_global_tau}--\eqref{eq:strict_global_nu};
 \State Construct the strict DD signatures using \eqref{eq:dd_signature};
 \State Recover the full-beam target-specific factor $\hat{\mathbf{A}}_{\theta}^{\mathrm{DD}}$ using \eqref{eq:A_theta_DD};
 \State Select beam index $\hat g_k$ for $k=1,\ldots,\hat K$ using \eqref{eq:g_k_DD};
 \Ensure $\hat K$, $\{\bar{\mathbf{a}}_{\tau,k},\bar{\mathbf{a}}_{\nu,k},\hat g_k\}_{k=1}^{\hat K}$, and $\hat{\mathbf{A}}_{\theta}^{\mathrm{DD}}$.
 \end{algorithmic}
\end{algorithm}

Algorithm~\ref{alg:global} summarizes the global processing stage.
The screened and compressed tensor provides robust DD separation, while \eqref{eq:A_theta_DD} restores target-specific responses over the complete set of beams.

%====================================================================
\section{Local Tensor Refinement and Parameter Estimation}
\label{sec:local_refinement}

The global stage provides strict DD factors and a target-specific beam response for each component.
The local stage then returns to the individual RF-chain observations at each selected beam and performs a high-resolution structured decomposition.
Under the adopted subarray geometry with $R\ge4$, the RF-chain spatial factor enables closed-form elevation and azimuth recovery from the subarray phase centers.

\subsection{Local Tensor Formulation}

The set of global components with the optimal beam being beam $g$ is defined as
\begin{equation}
 \mathcal{K}_{g}
 \triangleq
 \{k\mid\hat g_k=g,\;k=1,\ldots,\hat K\}.
\end{equation}
Let $I_g=|\mathcal{K}_g|$.
For each distinct beam with $I_g>0$, construct
$\mathcal{T}_g\in\mathbb{C}^{M_{\mathrm{c}}\times R\times Q}$ as
\begin{equation}\label{eq:local_tensor_definition}
 [\mathcal{T}_{g}]_{m,r,q}
 =
 [\hat{\mathbf{y}}_{m,g,q}]_r.
\end{equation}
Unlike the global tensor, $\mathcal{T}_g$ preserves the individual RF-chain responses.

Finite beamwidth and sidelobes can cause additional target components to appear in a local tensor even if their strongest selected beam is different from $g$.
The local rank is therefore estimated independently by the minimum description length (MDL) criterion \cite{wax1985}.
Let
$\mathbf{T}_{g,(2)}\in\mathbb{C}^{R\times M_{\mathrm{c}}Q}$
be the mode-2 unfolding and define
\begin{equation}\label{eq:local_covariance}
 \hat{\mathbf{R}}_{g}^{\mathrm{L}}
 =
 \frac{1}{M_{\mathrm{c}}Q}
 \mathbf{T}_{g,(2)}\mathbf{T}_{g,(2)}^{\mathrm{H}}.
\end{equation}
If
$\zeta_{g,1}^{\mathrm{L}}\ge\cdots\ge\zeta_{g,R}^{\mathrm{L}}\ge0$
are its eigenvalues, define $I_{\max}=\min\{K_{\max},R-1\}$. For $i\in\{0,\ldots,I_{\max}\}$,
\begin{align}
 \operatorname{MDL}_{g}(i)
 =
 &-M_{\mathrm{c}}Q(R-i)
 \ln\left(
 \frac{
 \left(\prod_{n=i+1}^{R}\zeta_{g,n}^{\mathrm{L}}\right)^{1/(R-i)}
 }{
 \dfrac{1}{R-i}\sum_{n=i+1}^{R}\zeta_{g,n}^{\mathrm{L}}
 }
 \right)
 \notag\\
 &+\frac{1}{2}i(2R-i)\ln(M_{\mathrm{c}}Q).
 \label{eq:local_MDL}
\end{align}
The local model order and decomposition rank are
\begin{align}
 \hat I_g
 &=
 \arg\min_{0\le i\le I_{\max}}\operatorname{MDL}_{g}(i),\\
 I_{g,\mathrm{deco}}
 &=
 \max\{I_g,\hat I_g\}.
 \label{eq:I_deco}
\end{align}

For notational simplicity, we consider one selected beam and omit the beam subscript $g$ for the remainder of this section.
The local tensor is approximated by the rank-$I_{\mathrm{deco}}$ CP model
\begin{equation}\label{eq:local_CPD}
 \mathcal{T}
 \approx
 \sum_{i=1}^{I_{\mathrm{deco}}}
 \alpha_i'
 \mathbf{c}_{\tau,i}
 \circ
 \mathbf{c}_{\mathrm{s},i}
 \circ
 \mathbf{c}_{\nu,i}
 +
 \mathcal{N}',
\end{equation}
where
\begin{align}
 \mathbf{c}_{\tau,i}
 &=
 \left[e^{-j2\pi m\Delta f\tau_i}\right]_{m=1}^{M_{\mathrm{c}}},
 \label{eq_c_tau_local}
 \\
 \mathbf{c}_{\mathrm{s},i}
 &=
 e^{j2\pi f_i^{\mathrm{d}}(g-1)T_{\mathrm{block}}}
 \mathbf{b}_{g,i},
 \label{eq_c_s}
 \\
 \mathbf{c}_{\nu,i}
 &=
 \left[e^{j2\pi f_i^{\mathrm{d}}(q-1)T_{\mathrm{s}}}\right]_{q=1}^{Q}.
 \label{eq_c_nu_local}
\end{align}
Here, $\alpha_i'$ absorbs the effective complex gain and the remaining CP scaling of local component $i$, and $\mathcal{N}'$ represents receiver noise and residual model mismatch. The spatial factor $\mathbf{c}_{\mathrm{s},i}$ contains the relative phases across the subarray phase centers that are required for angle estimation.

\subsection{Vandermonde-Constrained Rectified CP-ALS}

The local delay and intra-burst Doppler factors obey the same Vandermonde structures as their global counterparts.
We therefore employ a rectified ALS (RecALS) procedure \cite{boyer2016} in which each unconstrained delay and Doppler update is followed by a projection using the operator in \eqref{eq:strict_operator}.
Specifically, if
$\mathbf{c}_{\tau,i}^{\mathrm{ALS}}$
and
$\mathbf{c}_{\nu,i}^{\mathrm{ALS}}$
are the updated factor columns in one ALS iteration, they are replaced by
\begin{align}
 \mathbf{c}_{\tau,i}^{\mathrm{rec}}
 &=
 \mathcal{R}_{1}
 \left(\mathbf{c}_{\tau,i}^{\mathrm{ALS}}\right),
 \label{eq:delay_projection}
 \\
 \mathbf{c}_{\nu,i}^{\mathrm{rec}}
 &=
 \mathcal{R}_{0}
 \left(\mathbf{c}_{\nu,i}^{\mathrm{ALS}}\right).
 \label{eq:doppler_projection}
\end{align}
The spatial factor is updated by a regularized ALS subproblem and remains unconstrained. For numerical stability, the three local ALS subproblems use the same scale-adaptive diagonal Tikhonov form as the global update in \eqref{eq:reg_als_update}.
The ALS updates and Vandermonde projections are alternated until convergence.
After convergence, denote the resulting local factor vectors by
$\hat{\mathbf{c}}_{\tau,i}$,
$\hat{\mathbf{c}}_{\mathrm{s},i}$, and
$\hat{\mathbf{c}}_{\nu,i}$.
%where $\hat{\mathbf{c}}_{\tau,i}$ and $\hat{\mathbf{c}}_{\nu,i}$ are the final strict Vandermonde factors after rectification, while $\hat{\mathbf{c}}_{\mathrm{s},i}$ is the corresponding unconstrained RF-chain spatial factor.
Let $\hat\psi_{\tau,i}$ and $\hat\psi_{\nu,i}$ denote the phase increments obtained by the final applications of \eqref{eq:strict_coarse_slope}--\eqref{eq:strict_operator}.
The local delay estimate is
\begin{equation}\label{eq:tau_est}
 \hat\tau_i
 =
 \frac{
 \modulo(-\hat\psi_{\tau,i},2\pi)
 }{
 2\pi\Delta f
 },
\end{equation}
and the coarse intra-burst Doppler estimate is
\begin{equation}\label{eq:fd_coarse}
 \hat f_{i,\mathrm{coarse}}^{\mathrm{d}}
 =
 \frac{
 \mathcal{U}\!\left(e^{j\hat\psi_{\nu,i}}\right)
 }{
 2\pi T_{\mathrm{s}}
 }.
\end{equation}

\begin{remark}\label{remark:time_scale}
	\emph{
		The local intra-burst Doppler estimate in \eqref{eq:fd_coarse} has a wide unambiguous velocity range, nominally
		$|v|<\lambda/(4T_{\mathrm{s}})$, while its accuracy is limited by the short $Q$-symbol aperture.
		In contrast, the inter-burst Doppler phase in the global beam-domain factor provides a much longer temporal aperture and hence higher Doppler accuracy, at the cost of a narrower nominal unambiguous range,
		$|v|<\lambda/(4T_{\mathrm{block}})$.
		Therefore, the local intra-burst estimate serves as a coarse ambiguity-resolving reference, while the global inter-burst phase is used for fine residual-Doppler estimation.
		}
\end{remark}

\subsection{Subarray-Based Angle Estimation}

After the local CP decomposition, the RF-chain spatial responses of different targets are separated into individual columns.
Angle estimation can therefore be performed independently for each component.
Sum--difference processing has been investigated for angle estimation and receive-beam selection \cite{zhang2025sumdifference}.
We exploit the regular phase-center geometry of the subarray grid to derive exact sum--difference relations from finite geometric series, avoiding the small-angle approximation commonly used in monopulse formulas \cite{sherman2011}.

\subsubsection{Subarray Phase-Center Model}

Define the direction cosines of local component $i$ as
\begin{equation}
 u_i=\sin\theta_i\cos\phi_i,
 \quad
 u_{\mathrm{y},i}=\sin\theta_i\sin\phi_i,
 \quad
 w_i=\cos\theta_i.
\end{equation}
For the horizontal-first subarray ordering, the horizontal and vertical subarray indices of RF chain $r$ are
\begin{align}
 r_{\mathrm{h}}
 &=
 \modulo(r-1,2R^{\mathrm{h}})+1,\\
 r_{\mathrm{v}}
 &=
 \left\lfloor\frac{r-1}{2R^{\mathrm{h}}}\right\rfloor+1.
\end{align}
The spacings between adjacent subarray phase centers are
\begin{equation}
 D_{\mathrm{x}}=L^{\mathrm{sub,h}}d_{\mathrm{h}},
 \qquad
 D_{\mathrm{z}}=L^{\mathrm{sub,v}}d_{\mathrm{v}}.
\end{equation}
Due to the identical internal geometry of all subarrays that are steered toward the same scanning direction $(\theta_g^{\mathrm{scan}},\phi_g^{\mathrm{scan}})$,
%Because all subarrays have identical internal geometry and are steered toward the same scanning direction $(\theta_g^{\mathrm{scan}},\phi_g^{\mathrm{scan}})$, 
the $r$-th element of the local spatial factor satisfies
\begin{equation}\label{c_s_i_r}
 [\mathbf{c}_{\mathrm{s},i}]_r
 =
 \rho_i
 e^{j\kappa D_{\mathrm{x}}(r_{\mathrm{h}}-1)\Delta u_i}
 e^{j\kappa D_{\mathrm{z}}(r_{\mathrm{v}}-1)\Delta w_i},
\end{equation}
where the nonzero complex scalar $\rho_i$ collects all RF-chain-independent gain and phase terms, and the direction-cosine offsets are defined as
\begin{align}
 \Delta u_i
 &=
 \sin\theta_i\cos\phi_i
 -
 \sin\theta_g^{\mathrm{scan}}\cos\phi_g^{\mathrm{scan}},
 \label{Delta_u_i}
 \\
 \Delta w_i
 &=
 \cos\theta_i-\cos\theta_g^{\mathrm{scan}}.
 \label{Delta_w_i}
\end{align}

\subsubsection{Exact Sum--Difference Ratio}

Define
\begin{align}
 \mathbf{w}_{\Sigma}
 &=\mathbf{1}_{R},\\
 \mathbf{w}_{\Delta u}
 &=
 \mathbf{1}_{2R^{\mathrm{v}}}\otimes
 \left[\mathbf{1}_{R^{\mathrm{h}}}^{\mathrm{T}},-\mathbf{1}_{R^{\mathrm{h}}}^{\mathrm{T}}\right]^{\mathrm{T}},\\
 \mathbf{w}_{\Delta w}
 &=
 \left[\mathbf{1}_{R^{\mathrm{v}}}^{\mathrm{T}},-\mathbf{1}_{R^{\mathrm{v}}}^{\mathrm{T}}\right]^{\mathrm{T}}
 \otimes\mathbf{1}_{2R^{\mathrm{h}}}.
\end{align}
The corresponding combining outputs are
\begin{equation}\label{eq:comb_outputs}
 o_{\Sigma,i}=\mathbf{w}_{\Sigma}^{\mathrm{H}}\hat{\mathbf{c}}_{\mathrm{s},i},
 \quad
 o_{\Delta u,i}=\mathbf{w}_{\Delta u}^{\mathrm{H}}\hat{\mathbf{c}}_{\mathrm{s},i},
 \quad
 o_{\Delta w,i}=\mathbf{w}_{\Delta w}^{\mathrm{H}}\hat{\mathbf{c}}_{\mathrm{s},i}.
\end{equation}
Let
$\omega_{\mathrm{x},i}=\kappa D_{\mathrm{x}}\Delta u_i$
and
$\omega_{\mathrm{z},i}=\kappa D_{\mathrm{z}}\Delta w_i$.
In the noiseless case with an exactly recovered spatial factor, substituting \eqref{c_s_i_r} into \eqref{eq:comb_outputs} and evaluating the finite geometric series yields
\begin{align}
 \frac{o_{\Delta u,i}}{o_{\Sigma,i}}
 &=
 \frac{
 -j\,\dfrac{2\sin^2\!\left(R^{\mathrm{h}}\omega_{\mathrm{x},i}/2\right)}{\sin\!\left(\omega_{\mathrm{x},i}/2\right)}
 e^{j\omega_{\mathrm{x},i}(2R^{\mathrm{h}}-1)/2}
 }{
 \dfrac{\sin\!\left(R^{\mathrm{h}}\omega_{\mathrm{x},i}\right)}{\sin\!\left(\omega_{\mathrm{x},i}/2\right)}
 e^{j\omega_{\mathrm{x},i}(2R^{\mathrm{h}}-1)/2}
 }
 \notag\\
 &=
 -j\tan\left(\frac{R^{\mathrm{h}}\omega_{\mathrm{x},i}}{2}\right),
 \label{eq:ratio_u}
 \\
 \frac{o_{\Delta w,i}}{o_{\Sigma,i}}
 &=
 -j\tan\left(\frac{R^{\mathrm{v}}\omega_{\mathrm{z},i}}{2}\right).
 \label{eq:ratio_w}
\end{align}
These identities are exact under the adopted noiseless subarray model and require no small-angle approximation.

\subsubsection{Closed-Form Angle Recovery}

Within the principal intervals
\begin{equation}\label{eq:monopulse_interval}
 \left|\frac{R^{\mathrm{h}}\omega_{\mathrm{x},i}}{2}\right|<\frac{\pi}{2},
 \qquad
 \left|\frac{R^{\mathrm{v}}\omega_{\mathrm{z},i}}{2}\right|<\frac{\pi}{2},
\end{equation}
the tangent functions are one-to-one.
For the local component associated with the globally selected target, the target-specific optimal beam  selection in \eqref{eq:g_k_DD} ensures the component close to its scanning direction.
Therefore, the direction-cosine offsets can be estimated as
\begin{align}
	\widehat{\Delta u}_i
	&=
	\frac{2}{\kappa R^{\mathrm{h}}D_{\mathrm{x}}}
	\arctan\left(
	\Re\left\{j\frac{o_{\Delta u,i}}{o_{\Sigma,i}}\right\}
	\right),
	\label{eq:du_est}
	\\
	\widehat{\Delta w}_i
	&=
	\frac{2}{\kappa R^{\mathrm{v}}D_{\mathrm{z}}}
	\arctan\left(
	\Re\left\{j\frac{o_{\Delta w,i}}{o_{\Sigma,i}}\right\}
	\right).
	\label{eq:dw_est}
\end{align}
The direction cosines are then reconstructed as
\begin{align}
 \hat u_i
 &=
 \sin\theta_g^{\mathrm{scan}}\cos\phi_g^{\mathrm{scan}}
 +\widehat{\Delta u}_i,
 \label{eq:u_est}
 \\
 \hat w_i
 &=
 \cos\theta_g^{\mathrm{scan}}+\widehat{\Delta w}_i,
 \label{eq:w_est}
 \\
 \hat u_{\mathrm{y},i}
 &=
 \sqrt{\max\{0,1-\hat u_i^2-\hat w_i^2\}}.
 \label{eq:u_y_est}
\end{align}
The nonnegative square root in \eqref{eq:u_y_est} follows from the adopted azimuth interval $\phi\in[0,\pi)$, under which the $y$-axis direction cosine is nonnegative.
Finally,
\begin{align}
 \hat\theta_i
 &=
 \arccos\left(\max\{-1,\min\{1,\hat w_i\}\}\right),
 \label{eq:theta_est}
 \\
 \hat\phi_i
 &=
 \atanTwo(\hat u_{\mathrm{y},i},\hat u_i).
 \label{eq:phi_est}
\end{align}
In the noiseless case, this estimator exactly recovers $\Delta u_i$ and $\Delta w_i$ within the principal intervals in \eqref{eq:monopulse_interval}, without small-angle approximation bias.

\begin{algorithm}[t]
 \caption{Local Structured Tensor Decomposition and Parameter Extraction}
 \label{alg:local}
 \begin{algorithmic}[1]
 \Require Selected beam $g$, local observations $\{\hat{\mathbf{y}}_{m,g,q}\}$, and global component set $\mathcal{K}_g$.
 \State Construct $\mathcal{T}_g$ using \eqref{eq:local_tensor_definition};
 \State Estimate $\hat I_g$ by MDL and set $I_{g,\mathrm{deco}}=\max\{I_g,\hat I_g\}$;
 \State Initialize the rank-$I_{g,\mathrm{deco}}$ CP factors;
 \Repeat
 \State Update the delay factor by regularized ALS and rectify every column using $\mathcal{R}_1(\cdot)$;
 \State Update the RF-chain spatial factor by regularized ALS;
 \State Update the intra-burst Doppler factor by regularized ALS and rectify every column using $\mathcal{R}_0(\cdot)$;
 \Until{convergence}
 \For{$i=1,\ldots,I_{g,\mathrm{deco}}$}
 \State Obtain $\hat\tau_i$ and $\hat f_{i,\mathrm{coarse}}^{\mathrm{d}}$ using \eqref{eq:tau_est} and \eqref{eq:fd_coarse};
 \State Compute $o_{\Sigma,i}$, $o_{\Delta u,i}$, and $o_{\Delta w,i}$;
 \State Recover $\hat\theta_i$ and $\hat\phi_i$ using \eqref{eq:du_est}--\eqref{eq:phi_est};
 \EndFor
 \Ensure $\{\hat\tau_i,\hat f_{i,\mathrm{coarse}}^{\mathrm{d}},\hat\theta_i,\hat\phi_i\}_{i=1}^{I_{g,\mathrm{deco}}}$ and the strict local DD factors $\{\hat{\mathbf{c}}_{\tau,i},\hat{\mathbf{c}}_{\nu,i}\}_{i=1}^{I_{g,\mathrm{deco}}}$.
 \end{algorithmic}
\end{algorithm}

Algorithm~\ref{alg:local} summarizes the local structured decomposition and physical-parameter extraction.

%====================================================================
\section{Cross-Stage Association and Doppler Refinement}
\label{sec:association_refinement}

The global and local tensor decompositions are performed independently and therefore generally return target components in different orders.
Hence, cross-stage association is required. In addition, the local decomposition may also contain additional candidates due to the over-factoring in \eqref{eq:I_deco}.
In this section, we match the local candidates to the global components utilizing their common strict DD structures, and then use the associated local Doppler estimate as a coarse ambiguity-resolving reference for fine inter-burst Doppler refinement.

\subsection{Strict-DD-Based Cross-Stage Association}

Let
$\mathcal{I}_g=\{1,\ldots,I_{g,\mathrm{deco}}\}$
denote the local candidate set for selected beam $g$.
For global component $k\in\mathcal{K}_g$ and local candidate $i\in\mathcal{I}_g$, define the DD matching measure as
\begin{equation}\label{eq:matching}
 M_{k,i}
 =
 \frac{
 \left|\bar{\mathbf{a}}_{\tau,k}^{\mathrm{H}}\hat{\mathbf{c}}_{\tau,i}\right|
 }{
 \|\bar{\mathbf{a}}_{\tau,k}\|_2\|\hat{\mathbf{c}}_{\tau,i}\|_2
 }
 +
 \frac{
 \left|\bar{\mathbf{a}}_{\nu,k}^{\mathrm{H}}\hat{\mathbf{c}}_{\nu,i}\right|
 }{
 \|\bar{\mathbf{a}}_{\nu,k}\|_2\|\hat{\mathbf{c}}_{\nu,i}\|_2
 }.
\end{equation}
Note that both the global and local DD factors in \eqref{eq:matching} are strict Vandermonde vectors. The normalized inner products provide bounded and scale-invariant similarity measures for cross-stage association, with each term within the $[0,1]$ range.
To obtain a one-to-one association between the $|\mathcal{K}_g|$ global components and a subset of the $I_{g,\mathrm{deco}}$ local candidates, the following linear-assignment problem is formulated
\begin{align}\label{eq:association_problem}
 \max_{\{x_{k,i}\}}
 &\quad
 \sum_{k\in\mathcal{K}_{g}}
 \sum_{i\in\mathcal{I}_{g}}
 x_{k,i}M_{k,i}
 \\
 \mathrm{s.t.}
 &\quad
 \sum_{i\in\mathcal{I}_{g}}x_{k,i}=1,
 \quad \forall k\in\mathcal{K}_{g},
 \notag\\
 &\quad
 \sum_{k\in\mathcal{K}_{g}}x_{k,i}\le1,
 \quad \forall i\in\mathcal{I}_{g},
 \notag\\
 &\quad
 x_{k,i}\in\{0,1\}.
 \notag
\end{align}
This problem can be solved efficiently by the Hungarian algorithm \cite{kuhn1955}.

For feasibility screening, the delay and coarse Doppler estimates of local candidate $i$ are firstly converted to the range and radial velocity estimates
\begin{equation}
 \hat r_i^{\mathrm{L}}=\frac{c_0\hat\tau_i}{2},
 \qquad
 \hat v_{i,\mathrm{coarse}}^{\mathrm{L}}
 =\frac{\lambda\hat f_{i,\mathrm{coarse}}^{\mathrm{d}}}{2}.
\end{equation}
Let
$\mathcal{P}=[r_{\min},r_{\max}]\times[v_{\min},v_{\max}]\times[\theta_{\min},\theta_{\max}]\times[\phi_{\min},\phi_{\max}]$
denote the prescribed sensing region. After the similarity-based DD association, a matched local candidate is rejected if
$(\hat r_i^{\mathrm{L}},\hat v_{i,\mathrm{coarse}}^{\mathrm{L}},\hat\theta_i,\hat\phi_i)\notin\mathcal{P}$.
Based on the DD matching measures, the invalid matches can be replaced by the unused feasible local candidates generated by the over-factoring. This feasibility check suppresses spurious local components without modifying the DD-based association metric.

After the association and feasibility verification of all selected beams are completed, the accepted local estimates are reindexed. The resulting $\hat K$ associated parameter tuples are denoted by
\begin{equation}\label{eq:associated_local_estimates}
 \left\{
 \left(
 \hat\tau_k,
 \hat f_{k,\mathrm{coarse}}^{\mathrm{d}},
 \hat\theta_k,
 \hat\phi_k
 \right)
 \right\}_{k=1}^{\hat K},
\end{equation}
where the index $k$ follows the global-stage component ordering.

\subsection{Spatial-Mask-Aided WLS Doppler Refinement}

To perform WLS Doppler refinement, denote the \(k\)th column of $\hat{\mathbf A}_{\theta}^{\mathrm{DD}}$ by $\hat{\mathbf a}_{\theta,k}^{\mathrm{DD}}\in\mathbb C^{G\times 1}$, whose $g$-th entry $\hat a_{\theta,k}^{\mathrm{DD}}(g)$ represents the recovered complex response of target $k$ under beam $g$.
From \eqref{eq:d_g}, the $g$-th entry of the reconstructed full-beam factor satisfies
\begin{equation}\label{eq:ddrp_interburst_model}
 \hat a_{\theta,k}^{\mathrm{DD}}(g)
 \approx
 \alpha_k\beta_{g,k}
 e^{j2\pi f_k^{\mathrm{d}}t_g},
 \qquad
 t_g\triangleq t_{g,1}=(g-1)T_{\mathrm{block}}.
\end{equation}
Since $\beta_{g,k}$ is real and nonnegative, the inter-burst phase progression in \eqref{eq:ddrp_interburst_model} is determined by the Doppler frequency, apart from the beam-independent phase of $\alpha_k$. However, the phase becomes unreliable on beams with very weak target responses. We therefore retain only a spatial neighborhood around the reselected beam and weight its samples according to the beam gain predicted from the locally refined angle.

Specifically, let $d_{\mathrm{nb}}\ge0$ denote the neighborhood radius
and $\varepsilon_{\mathrm{num}}>0$ a numerical tolerance. Define
\begin{equation}\label{eq:Gk_neighborhood}
	\mathcal{G}_k
	=
	\left\{
	g\in\{1,\ldots,G\}
	\;\middle|\;
	\begin{aligned}
		&\|\boldsymbol{\pi}_g-\boldsymbol{\pi}_{\hat g_k}\|_{\infty}
		\le d_{\mathrm{nb}},\\
		&|\hat a_{\theta,k}^{\mathrm{DD}}(g)|>\varepsilon_{\mathrm{num}}
	\end{aligned}
	\right\}.
\end{equation}
For $\hat{\mathbf{a}}_k=\mathbf{a}(\hat\theta_k,\hat\phi_k)$, we define the corresponding theoretical beam gain  as
\begin{align}
 \beta_{g,k}^{\mathrm{th}}
 =
 \left|
 \mathbf{1}_{R}^{\mathrm{H}}
 \left(\mathbf{W}_{g}^{\Sigma}\right)^{\mathrm{H}}
 \hat{\mathbf{a}}_k
 \right|^2.
 \label{eq:beta_th}
\end{align}
Under the approximately beam-independent DDRP noise, the phase-error variance is inversely proportional to the squared beam response at a sufficiently high signal-to-noise ratio.
Hence, the squared relative gain provides the inverse-variance weighting for the phase-domain WLS estimator:
\begin{align}
 \omega_{g,k}
 =
 \left(
 \frac{\beta_{g,k}^{\mathrm{th}}}
 {\max_{g'\in\mathcal{G}_k}\beta_{g',k}^{\mathrm{th}}}
 \right)^2,
 \qquad g\in\mathcal{G}_k.
 \label{eq:doppler_weight}
\end{align}

 \begin{remark}\label{remark:spatial_mask}
 	\emph{Although DDRP provides target-specific responses over all scanned beams, phases from beams far from the target direction can be noise dominated. Restricting the WLS fit to $\mathcal{G}_k$ sacrifices part of the inter-burst temporal aperture in exchange for more reliable phase samples. The neighborhood radius $d_{\mathrm{nb}}$ controls this trade-off.}
 \end{remark}

Using the intra-burst Doppler estimate $\hat f_{k,\mathrm{coarse}}^{\mathrm{d}}$ with a wide unambiguous range, we firstly remove the coarse Doppler phase from the DDRP factor:
\begin{align}
 \tilde x_{g,k}
 &=
 \hat a_{\theta,k}^{\mathrm{DD}}(g)
 e^{-j2\pi\hat f_{k,\mathrm{coarse}}^{\mathrm{d}}t_g}
 \notag\\
 &\approx
 \alpha_k\beta_{g,k}
 e^{j2\pi\Delta f_k^{\mathrm{d}}t_g},
 \qquad g\in\mathcal{G}_k,
 \label{eq:interburst_derotation}
\end{align}
where
$\Delta f_k^{\mathrm{d}}=f_k^{\mathrm{d}}-\hat f_{k,\mathrm{coarse}}^{\mathrm{d}}$
is the residual Doppler frequency. For the beams in $\mathcal{G}_k$, let $(g_1,\ldots,g_{|\mathcal{G}_k|})$ denote their indices ordered by increasing time stamp $t_g$. 
%A sufficient noiseless condition for correct phase unwrapping is
%\begin{equation*}
% |\Delta f_k^{\mathrm{d}}|(t_{g_j}-t_{g_{j-1}})<\frac{1}{2},
% \quad j=2,\ldots,|\mathcal{G}_k|.
%\end{equation*}
The residual phases are obtained from the complex observations as
\begin{align}
 \varphi_{g_1,k}
 &=
 \mathcal{U}(\tilde x_{g_1,k}),\notag\\
 \varphi_{g_j,k}
 &=
 \varphi_{g_{j-1},k}
 +
 \mathcal{U}\!\left(\tilde x_{g_{j-1},k}^{*}\tilde x_{g_j,k}\right),
 \, j=2,\ldots,|\mathcal{G}_k|.
 \label{eq:interburst_phase_unwrap}
\end{align}
The resulting unwrapped phase follows the approximately linear model
\begin{equation}\label{eq:residual_phase_model}
 \varphi_{g,k}
 \approx
 \varphi_{0,k}
 +2\pi\Delta f_k^{\mathrm{d}}t_g,
 \qquad g\in\mathcal{G}_k,
\end{equation}
where $\varphi_{0,k}$ is an unknown constant phase.
Therefore, the residual Doppler can be estimated by weighted linear regression. Define
\begin{equation}
 \bar t_k
 =
 \frac{\sum_{g\in\mathcal{G}_k}\omega_{g,k}t_g}
 {\sum_{g\in\mathcal{G}_k}\omega_{g,k}},
 \qquad
 \bar\varphi_k
 =
 \frac{\sum_{g\in\mathcal{G}_k}\omega_{g,k}\varphi_{g,k}}
 {\sum_{g\in\mathcal{G}_k}\omega_{g,k}}.
\end{equation}
The WLS slope gives
\begin{equation}\label{eq:fine_doppler}
 \Delta\hat f_k^{\mathrm{d}}
 =
 \frac{1}{2\pi}
 \frac{
 \sum_{g\in\mathcal{G}_k}
 \omega_{g,k}(t_g-\bar t_k)(\varphi_{g,k}-\bar\varphi_k)
 }{
 \sum_{g\in\mathcal{G}_k}
 \omega_{g,k}(t_g-\bar t_k)^2
 }.
\end{equation}
%If the weighted time variance in \eqref{eq:fine_doppler} vanishes, we set $\Delta\hat f_k^{\mathrm{d}}=0$ and retain the coarse estimate.
Finally, the refined range and radical velocity estimation parameters are obtained as
\begin{align}
 \hat f_k^{\mathrm{d}}
 &=
 \hat f_{k,\mathrm{coarse}}^{\mathrm{d}}
 +\Delta\hat f_k^{\mathrm{d}},
 \label{eq:final_doppler}
 \\
 \hat r_k
 &=
 \frac{c_0\hat\tau_k}{2},
 \qquad
 \hat v_k
 =
 \frac{\lambda\hat f_k^{\mathrm{d}}}{2}.
 \label{eq:range_velocity_final}
\end{align}

\begin{algorithm}[t]
 \caption{Overall Structured Dual-Tensor Sensing Scheme}
 \label{alg:overall}
 \begin{algorithmic}[1]
 \Require Demodulated observations $\hat{\mathbf{y}}_{m,g,q}$, scan-grid locations $\{\boldsymbol{\pi}_g\}_{g=1}^{G}$, and sensing region $\mathcal{P}$.
 \State Execute Algorithm~\ref{alg:global};
 \For{each distinct selected beam $g$}
 \State Construct $\mathcal{K}_g$ and execute Algorithm~\ref{alg:local};
 \State Compute the DD matching measures using \eqref{eq:matching};
 \State Associate local candidates by solving \eqref{eq:association_problem}, apply the sensing-region feasibility check, and use an unused feasible candidate as a replacement when needed;
 \EndFor
 \State Reindex the associated local estimates according to \eqref{eq:associated_local_estimates};
 \For{$k=1,\ldots,\hat K$}
 \State Construct $\mathcal{G}_k$ and $\omega_{g,k}$ using \eqref{eq:Gk_neighborhood}--\eqref{eq:doppler_weight};
 \State Obtain $\Delta\hat f_k^{\mathrm{d}}$ via \eqref{eq:fine_doppler};
 \State Compute $\hat f_k^{\mathrm{d}}$, $\hat r_k$, and $\hat v_k$ using \eqref{eq:final_doppler}--\eqref{eq:range_velocity_final};
 \EndFor
 \Ensure Final parameter estimates $(\hat r_k,\hat v_k,\hat\theta_k,\hat\phi_k)$ for all resolved targets.
 \end{algorithmic}
\end{algorithm}

Algorithm~\ref{alg:overall} summarizes the complete sensing procedure.

\subsection{Computational Complexity}

The computational cost is summarized according to the three main processing blocks. Let $N_{\mathrm{iter}}^{\mathrm{G}}$ and $N_{\mathrm{iter}}^{\mathrm{L}}$ denote the number of global and local ALS iterations, respectively.

%\textit{First-stage tensor decomposition.}
Firstly, the global stage firstly performs a rank-$R_1$ mode-1 truncation on the screened tensor and then applies rank-$\hat K$ regularized CP-ALS to the reduced tensor of size $R_1\times G_{\mathrm{keep}}\times Q$.
Using a truncated singular value decomposition, the corresponding complexity is
\begin{align}
 C_{\mathrm{G,CP}}
 =\mathcal{O}\Big(&
 M_{\mathrm{c}}G_{\mathrm{keep}}QR_1
 +N_{\mathrm{iter}}^{\mathrm{G}}
 \big[
 \hat K R_1G_{\mathrm{keep}}Q
 \notag\\
 &+\hat K^2(R_1+G_{\mathrm{keep}}+Q)
 +\hat K^3
 \big]
 \Big).
 \label{eq:complexity_global_cp}
\end{align}
The subsequent strict Vandermonde rectification has only linear complexity in the factor dimensions and is negligible in comparison.

%\textit{DDRP.}
Secondly, the DDRP scheme constructs the DD Gram matrix, solves the $\hat K\times\hat K$ regularized system, and projects the original observations over all $G$ beams, requiring
\begin{equation}
 C_{\mathrm{DDRP}}
 =
 \mathcal{O}\!\left(
 GM_{\mathrm{c}}Q\hat K
 +\hat K^2M_{\mathrm{c}}Q
 +\hat K^3
 \right).
 \label{eq:complexity_ddrp}
\end{equation}
This is a one-shot reprojection and involves no iterative tensor decomposition.

%\textit{Second-stage local refinement.}
Finally, the local RecALS decompositions on tensors of size $M_{\mathrm{c}}\times R\times Q$ have total complexity
\begin{align}
 C_{\mathrm{L}}
 =\mathcal{O}\Big(&
 N_{\mathrm{beam}}^{\mathrm{loc}}N_{\mathrm{iter}}^{\mathrm{L}}
 \big[
 I_{\mathrm{L}}M_{\mathrm{c}}RQ
 \notag\\
 &+I_{\mathrm{L}}^2(M_{\mathrm{c}}+R+Q)
 +I_{\mathrm{L}}^3
 \big]
 \Big),
 \label{eq:complexity_local}
\end{align}
where $N_{\mathrm{beam}}^{\mathrm{loc}}\le\hat K$ denotes the number of distinct selected beams, and  $I_{\mathrm{L}}=\max_g I_{g,\mathrm{deco}}$.
The local MDL estimation, cross-stage association, closed-form angle recovery, and WLS Doppler refinement incur comparatively small additional costs.

For the intended few-RF-chain configuration, $R$ and the number of selected local beams are both small. Thus, the overall computational burden is dominated by the first-stage global tensor processing, particularly the mode reduction and iterative CP decomposition, while DDRP and the second-stage refinements add substantially lower computational overhead.

%====================================================================
\section{Multi-Target CRLB Analysis}
\label{sec:crlb}

In this section, we derive the deterministic multi-target Cram\'{e}r--Rao lower bound (CRLB) for jointly estimating the range, radial velocity, elevation, and azimuth of all targets. Unlike target-wise single-target bounds, the joint formulation retains the coupling among nonorthogonal target signatures in delay, Doppler, and angle. The derivation follows the Slepian--Bangs formula for complex Gaussian observations \cite{kay1993}.

\subsection{Observation Model and Equivalent Fisher Information Matrix}

For target $k$, define the physical parameter vector as
\begin{equation}
 \boldsymbol{\xi}_k
 =
 [r_k,v_k,\theta_k,\phi_k]^{\mathrm{T}}
 \in\mathbb{R}^{4},
\end{equation}
and collect all targets as
\begin{equation}
 \boldsymbol{\xi}
 =
 [\boldsymbol{\xi}_1^{\mathrm{T}},\ldots,\boldsymbol{\xi}_K^{\mathrm{T}}]^{\mathrm{T}}
 \in\mathbb{R}^{4K}.
\end{equation}
Let $\boldsymbol{\alpha}=[\alpha_1,\ldots,\alpha_K]^{\mathrm{T}}$ contain the unknown deterministic complex gains, which are treated as nuisance parameters.

Using $\tau_k=2r_k/c_0$ and $f_k^{\mathrm{d}}=2v_k/\lambda$, the gain-free RF-chain response of target $k$ at sample $(m,g,q)$ is defined as
\begin{equation}\label{eq:gain_free_multitarget}
 \mathbf{s}_{m,g,q,k}
 =
 \mathbf{b}_{g,k}
 e^{-j4\pi m\Delta f r_k/c_0}
 e^{j4\pi v_k t_{g,q}/\lambda}.
\end{equation}
Accordingly, the mean of the demodulated observation in \eqref{eq:yhat_expanded} is
\begin{equation}\label{eq:multi_mean_sample}
 \boldsymbol{\mu}_{m,g,q}
 =
 \sum_{k=1}^{K}
 \alpha_k\mathbf{s}_{m,g,q,k}.
\end{equation}
Stacking all $M_{\mathrm{c}}GQ$ RF-chain observations gives
\begin{equation}\label{eq:stacked_multi_mean}
 \boldsymbol{\mu}
 =
 \sum_{k=1}^{K}\alpha_k\mathbf{s}_k
 =
 \mathbf{S}\boldsymbol{\alpha},
\end{equation}
where $\mathbf{s}_k\in\mathbb{C}^{M_{\mathrm{c}}GQR}$ stacks $\mathbf{s}_{m,g,q,k}$ and $\mathbf{S}=[\mathbf{s}_1,\ldots,\mathbf{s}_K]$.

Assuming independent receiver noise across $(m,g,q)$, the covariance of the stacked noise vector is
\begin{equation}\label{eq:stacked_noise_cov}
 \mathbf{C}
 =
 \operatorname{blkdiag}_{m,g,q}
 \left\{\mathbf{C}_{\mathrm{n},g}\right\}.
\end{equation}
If the antenna-domain receiver noise is spatially white with covariance $\sigma_{\mathrm{n}}^2\mathbf{I}_{L}$, then the adopted disjoint-subarray beamformer gives
\begin{equation}
 \mathbf{C}_{\mathrm{n},g}
 =
 \sigma_{\mathrm{n}}^2
 \left(\mathbf{W}_{g}^{\Sigma}\right)^{\mathrm{H}}
 \mathbf{W}_{g}^{\Sigma}
 =
 \frac{\sigma_{\mathrm{n}}^2}{R}\mathbf{I}_{R},
\end{equation}
and hence $\mathbf{C}=(\sigma_{\mathrm{n}}^2/R)\mathbf{I}_{M_{\mathrm{c}}GQR}$. The derivation below also applies to a known coloured covariance by retaining $\mathbf{C}$ explicitly.

For target $k$, define the Jacobian block of the stacked mean with respect to its physical parameters as
\begin{equation}\label{eq:D_xi_k}
 \mathbf{D}_{\xi,k}
 =
 \left[
 \frac{\partial\boldsymbol{\mu}}{\partial r_k},
 \frac{\partial\boldsymbol{\mu}}{\partial v_k},
 \frac{\partial\boldsymbol{\mu}}{\partial\theta_k},
 \frac{\partial\boldsymbol{\mu}}{\partial\phi_k}
 \right]
 \in\mathbb{C}^{M_{\mathrm{c}}GQR\times4},
\end{equation}
and let $\mathbf{D}_{\xi}=[\mathbf{D}_{\xi,1},\ldots,\mathbf{D}_{\xi,K}]$. To eliminate the unknown complex gains, whiten the response and Jacobian matrices as
\begin{equation}
 \bar{\mathbf{S}}
 =
 \mathbf{C}^{-1/2}\mathbf{S},
 \qquad
 \bar{\mathbf{D}}_{\xi}
 =
 \mathbf{C}^{-1/2}\mathbf{D}_{\xi},
\end{equation}
and define the orthogonal projector onto the complement of the gain subspace as
\begin{equation}\label{eq:gain_projector}
 \boldsymbol{\Pi}_{\bar{S}}^{\perp}
 =
 \mathbf{I}
 -
 \bar{\mathbf{S}}
 (\bar{\mathbf{S}}^{\mathrm{H}}\bar{\mathbf{S}})^{\dagger}
 \bar{\mathbf{S}}^{\mathrm{H}},
\end{equation}
where $\dagger$ denotes the Moore--Penrose pseudoinverse. The equivalent Fisher information matrix (EFIM) for the $4K$ physical parameters is given by
\begin{equation}\label{eq:multi_EFIM}
 \mathbf{F}_{\mathrm{eq}}
 =
 2\Re\left\{
 \bar{\mathbf{D}}_{\xi}^{\mathrm{H}}
 \boldsymbol{\Pi}_{\bar{S}}^{\perp}
 \bar{\mathbf{D}}_{\xi}
 \right\}
 \in\mathbb{R}^{4K\times4K}.
\end{equation}
This form is equivalent to eliminating the real and imaginary parts of $\boldsymbol{\alpha}$ from the full Slepian--Bangs Fisher information matrix by a Schur complement.

Let $\bar{\mathbf{D}}_{\xi,k}=\mathbf{C}^{-1/2}\mathbf{D}_{\xi,k}$. The $(k,\ell)$-th block of \eqref{eq:multi_EFIM} is
\begin{equation}\label{eq:cross_target_EFIM}
 [\mathbf{F}_{\mathrm{eq}}]_{k,\ell}
 =
 2\Re\left\{
 \bar{\mathbf{D}}_{\xi,k}^{\mathrm{H}}
 \boldsymbol{\Pi}_{\bar{S}}^{\perp}
 \bar{\mathbf{D}}_{\xi,\ell}
 \right\}
 \in\mathbb{R}^{4\times4},
\end{equation}
which retains the cross-information between different targets. When the multi-target model is locally identifiable, the joint CRLB for locally unbiased estimation is
\begin{equation}\label{eq:multi_CRLB}
 \mathbf{C}_{\xi}^{\mathrm{CRLB}}
 =
 \mathbf{F}_{\mathrm{eq}}^{-1}.
\end{equation}
Let $[\mathbf{C}_{\xi}^{\mathrm{CRLB}}]_{k,k}\in\mathbb{R}^{4\times4}$ denote the $k$-th diagonal block, whose diagonal entries give the variance lower bounds for $r_k$, $v_k$, $\theta_k$, and $\phi_k$, respectively. When the cross-target blocks in \eqref{eq:cross_target_EFIM} are negligible, the joint bound approaches the corresponding independent single-target bounds. 
%If two target signatures become locally indistinguishable, $\mathbf{F}_{\mathrm{eq}}$ becomes singular and finite CRLBs for the individually labeled parameters do not exist.

\subsection{Jacobian Derivatives}

From \eqref{eq:gain_free_multitarget}--\eqref{eq:multi_mean_sample}, the range and radial velocity derivatives of target $k$ are
\begin{align}
 \frac{\partial\boldsymbol{\mu}_{m,g,q}}{\partial r_k}
 &=
 -j\frac{4\pi m\Delta f}{c_0}
 \alpha_k\mathbf{s}_{m,g,q,k},
 \\
 \frac{\partial\boldsymbol{\mu}_{m,g,q}}{\partial v_k}
 &=
 j\frac{4\pi t_{g,q}}{\lambda}
 \alpha_k\mathbf{s}_{m,g,q,k}.
\end{align}

Let $\mathbf{x}_{\mathrm{a}},\mathbf{z}_{\mathrm{a}}\in\mathbb{R}^{L\times  1}$ denote the antenna-coordinate vectors, where $[\mathbf{x}_{\mathrm{a}}]_l$ and $[\mathbf{z}_{\mathrm{a}}]_l$ are the horizontal and vertical coordinates, respectively, of the antenna element corresponding to the $l$-th entry of the steering vector. With $\mathbf{a}_k=\mathbf{a}(\theta_k,\phi_k)$, define
\begin{align}
 \mathbf{D}_{\theta,k}^{\mathrm{a}}
 &=
 \diag\left(
 j\kappa
 [\mathbf{x}_{\mathrm{a}}\cos\theta_k\cos\phi_k
 -\mathbf{z}_{\mathrm{a}}\sin\theta_k]
 \right),
 \\
 \mathbf{D}_{\phi,k}^{\mathrm{a}}
 &=
 \diag\left(
 -j\kappa\mathbf{x}_{\mathrm{a}}
 \sin\theta_k\sin\phi_k
 \right),
\end{align}
so that
\begin{equation}
 \dot{\mathbf{a}}_k^{(\theta)}
 =
 \mathbf{D}_{\theta,k}^{\mathrm{a}}\mathbf{a}_k,
 \qquad
 \dot{\mathbf{a}}_k^{(\phi)}
 =
 \mathbf{D}_{\phi,k}^{\mathrm{a}}\mathbf{a}_k.
\end{equation}
For $\chi\in\{\theta,\phi\}$, differentiating the RF-chain response in \eqref{b_gk} gives
\begin{equation}\label{eq:b_angle_derivative}
 \dot{\mathbf{b}}_{g,k}^{(\chi)}
 =
\sqrt{\frac{P_{\mathrm{t}}}{M_{\mathrm{c}}}}
 \left(\mathbf{W}_{g}^{\Sigma}\right)^{\mathrm{H}}
 \left[
 \dot{\mathbf{a}}_k^{(\chi)}\mathbf{a}_k^{\mathrm{H}}
 +
 \mathbf{a}_k(\dot{\mathbf{a}}_k^{(\chi)})^{\mathrm{H}}
 \right]
 \mathbf{W}_{g}^{\Sigma}\mathbf{1}_{R}.
\end{equation}
Hence,
\begin{align}\label{eq:angular_mean_derivative}
 \frac{\partial\boldsymbol{\mu}_{m,g,q}}{\partial\chi_k}
 &=
 \alpha_k
 \dot{\mathbf{b}}_{g,k}^{(\chi)}
 e^{-j4\pi m\Delta f r_k/c_0}
 e^{j4\pi v_k t_{g,q}/\lambda},
 \quad
 \chi\in\{\theta,\phi\}.
\end{align}
Stacking these sample-wise derivatives yields $\mathbf{D}_{\xi,k}$ in \eqref{eq:D_xi_k}.

\subsection{3D Positioning Error Bound}

Let $\mathbf{T}_{r\theta\phi}\in\mathbb{R}^{3K\times4K}$ denote the selection matrix that extracts
the range, elevation, and azimuth parameters of all targets from
$\boldsymbol{\xi}$, i.e.,
$\mathbf{T}_{r\theta\phi}\boldsymbol{\xi}
=
[r_1,\theta_1,\phi_1,\ldots,r_K,\theta_K,\phi_K]^{\mathrm{T}}$.
The Cartesian position of target $k$ is
\begin{equation}
 \mathbf{p}_k
 =
 \begin{bmatrix}
 r_k\sin\theta_k\cos\phi_k\\
 r_k\sin\theta_k\sin\phi_k\\
 r_k\cos\theta_k
 \end{bmatrix},
\end{equation}
with spherical-to-Cartesian Jacobian
\begin{align}
 \setlength{\arraycolsep}{2pt}
 \mathbf{J}_{\mathrm{pos},k}
 =
 \begin{bmatrix}
 \sin\theta_k\cos\phi_k
 &r_k\cos\theta_k\cos\phi_k
 &-r_k\sin\theta_k\sin\phi_k\\
 \sin\theta_k\sin\phi_k
 &r_k\cos\theta_k\sin\phi_k
 &r_k\sin\theta_k\cos\phi_k\\
 \cos\theta_k
 &-r_k\sin\theta_k
 &0
 \end{bmatrix}.
\end{align}
Define
\begin{equation}
 \mathbf{J}_{\mathrm{pos}}
 =
 \operatorname{blkdiag}
 \left\{
 \mathbf{J}_{\mathrm{pos},1},\ldots,
 \mathbf{J}_{\mathrm{pos},K}
 \right\}.
\end{equation}
The joint Cartesian position CRLB is therefore
\begin{equation}\label{eq:joint_position_crlb}
 \mathbf{C}_{\mathbf{p}}^{\mathrm{CRLB}}
 =
 \mathbf{J}_{\mathrm{pos}}
 \mathbf{T}_{r\theta\phi}
 \mathbf{C}_{\xi}^{\mathrm{CRLB}}
 \mathbf{T}_{r\theta\phi}^{\mathrm{T}}
 \mathbf{J}_{\mathrm{pos}}^{\mathrm{T}}
 \in\mathbb{R}^{3K\times3K}.
\end{equation}
Let $[\mathbf{C}_{\mathbf{p}}^{\mathrm{CRLB}}]_{k,k}$ denote its $k$-th diagonal $3\times3$ block. The 3D positioning root-mean-square error (RMSE) lower bound for target $k$ is
\begin{equation}\label{eq:position_rmse_lb}
 \operatorname{RMSE}_{\mathrm{3D,pos},k}^{\mathrm{lb}}
 =
 \sqrt{
 \Tr\left(
 [\mathbf{C}_{\mathbf{p}}^{\mathrm{CRLB}}]_{k,k}
 \right)
 }.
\end{equation}
A pooled per-target RMSE lower bound is
\begin{equation}\label{eq:average_position_rmse_lb}
 \operatorname{RMSE}_{\mathrm{3D,pos,avg}}^{\mathrm{lb}}
 =
 \sqrt{
 \frac{1}{K}
 \Tr\left(
 \mathbf{C}_{\mathbf{p}}^{\mathrm{CRLB}}
 \right)
 }.
\end{equation}
The bound assumes the correct model order and locally identifiable target labels under the deterministic point-target model in \eqref{H_mgq}. Model-order errors and association failures are not represented by this bound.

%====================================================================
\section{Simulation Results}
\label{sec:simulation}

In this section, representative numerical results are presented to evaluate the performance of the proposed beam-scanning-based MIMO-OFDM ISAC framework and structured dual-tensor sensing scheme.

\subsection{Simulation Setup}
\label{subsec:simulation_setup}

Unless otherwise stated, the BS employs a half-wavelength $16\times16$ UPA partitioned into a $2\times2$ subarray grid and connected to $R=4$ RF chains. The carrier frequency and subcarrier spacing are set to $f_{\mathrm{c}}=28$~GHz and $\Delta f=60$~kHz, respectively. Following the 50~MHz NR channel configuration \cite{3gpp38104}, 66 resource blocks (RBs) with $M_{\mathrm{c}}=792$ active subcarriers provide an allocated bandwidth of 47.52~MHz. The extended cyclic prefix is adopted to preserve uniform Doppler sampling \cite{3gpp38211}. The angular region $\theta\in[20^\circ,80^\circ]$ and $\phi\in[40^\circ,140^\circ]$ is covered by $G=16\times10=160$ scanning beams. Each burst contains $Q=10$ consecutive OFDM symbols using 4-QAM, and the inter-burst guard interval is set to $T_{\mathrm{gap}}=2T_{\mathrm{s}}$.

The targets are randomly and uniformly generated in the above spherical sector with ranges in $[10,100]$~m, subject to a minimum inter-target distance of 10~m. Their speeds are uniformly distributed over $[5,100]$~km/h with isotropically distributed motion directions, and the radar cross section is fixed at $0.05$~m$^2$. The noise power spectral density is $-174$~dBm/Hz. The target number and total transmit power are varied over $K\in\{1,2,3,4,5\}$ and $P_{\mathrm{t}}\in\{25,30,35,40,45,50\}$~dBm, respectively. The maximum number of global ALS iterations is $10^4$, and the convergence tolerance is $10^{-10}$.

For the $n$-th Monte Carlo trial, the scene-wise RMSE of $\chi\in\{r,v,\theta,\phi,\mathbf{p}\}$ is defined as
\begin{equation}
 e_{\chi}^{(n)}=\sqrt{\frac{1}{K}\sum_{k=1}^{K}
 \left\|\hat{\chi}_{k}^{(n)}-\chi_{k}^{(n)}\right\|_2^2}.
\end{equation}
Each RMSE result is obtained from $10^4$ independent trials and averaged over the 95\% of trials with the smallest scene-wise RMSEs. 
%All schemes use the same target realizations, transmitted symbols, and noise samples.

Three benchmarks are considered. \textbf{BasicBeam} directly uses the selected scanning-beam direction as the angle estimate. Estimation of signal parameters via rotational invariance techniques (\textbf{ESPRIT}) replaces the proposed global tensor-based DD extraction with its total least squares (TLS) variant \cite{roy1989}. Smoothed multiple signal classification (MUSIC) \cite{shan1985smoothing} is applied successively in the DD and angular domains to form the cascaded 2D MUSIC (\textbf{C2D-MUSIC}) benchmark. The fine search steps of C2D-MUSIC for target ranges, radial velocities, and angles are 0.005~m, 0.01~m/s, and $0.005^\circ$, respectively.
In addition, the multi-target CRLB in Section~\ref{sec:crlb} is used as the theoretical reference.

\subsection{Sensing Parameter Estimation Performance}
\label{subsec:parameter_estimation_results}

Firstly, we evaluate the accuracy of the information-theoretic-based AIC estimator for target-number estimation.
Table~\ref{tab:target_number_estimation} reports the probability of correctly estimating the target number using the estimator in \eqref{eq:AIC}.
All observed errors satisfy $|\hat K-K|\leq1$.
At a transmit power of 25~dBm, the probability of correct target-number estimation ranges from 99.43\% for $K=2$ to 96.11\% for $K=5$.
At 35~dBm, the probability reaches at least 99.25\% for all four target numbers.
Thus, the AIC estimate seldom supplies an incorrect component count to the subsequent global decomposition and local beam selection, including in the more crowded scenes.

\begin{table}[!htbp]
 \centering
 \caption{Probability of correct target-number estimation (\%) with $R=4$.}
 \label{tab:target_number_estimation}
 \renewcommand{\arraystretch}{1.08}
 \setlength{\tabcolsep}{7pt}
 \begin{tabular}{c|cccc}
  \hline
  $P_{\mathrm{t}}$ (dBm) & $K=2$ & $K=3$ & $K=4$ & $K=5$ \\
  \hline
  25 & 99.43 & 98.88 & 97.85 & 96.11 \\
  30 & 99.61 & 99.43 & 99.09 & 98.47 \\
  35 & 99.69 & 99.67 & 99.51 & 99.25 \\
  40 & 99.72 & 99.73 & 99.66 & 99.55 \\
  45 & 99.71 & 99.76 & 99.75 & 99.67 \\
  50 & 99.71 & 99.76 & 99.76 & 99.70 \\
  \hline
 \end{tabular}
\end{table}

\begin{figure}[t]
 \centering
 \includegraphics[width=0.98\columnwidth]{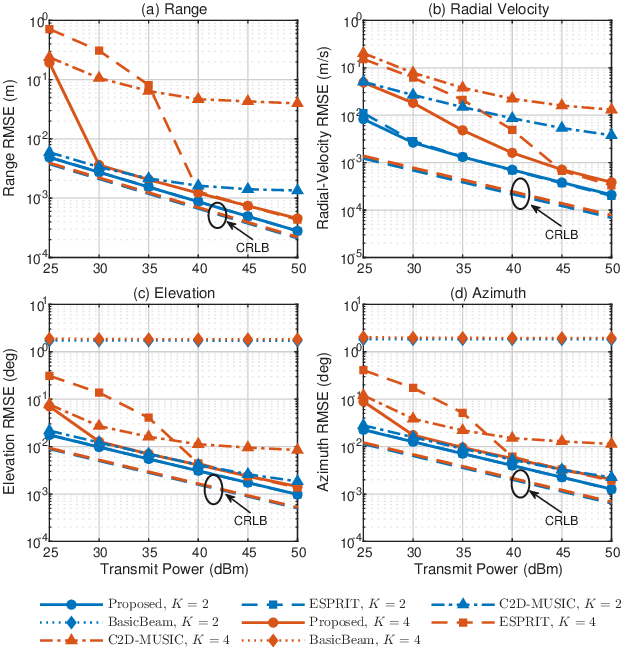}
 \caption{RMSEs of sensing-parameter estimates versus transmit power with $R=4$.}
 \label{fig:parameter_rmse_vs_power}
\end{figure}

Fig.~\ref{fig:parameter_rmse_vs_power} compares the four sensing-parameter RMSEs for $K=2$ and $K=4$. It can be seen from the figure that with two targets, the proposed range and radial-velocity errors decrease approximately in parallel with the CRLB, reaching submillimeter and submillimeter-per-second accuracy at high powers, while the angular errors fall to the millidegree level. Unlike BasicBeam, whose angular estimates are confined to the scan grid, the proposed local processing recovers continuous angle offsets from the separated RF-chain responses, yielding substantially lower angular RMSEs at high powers.
With four targets, the difference from ESPRIT and C2D-MUSIC is most pronounced around 35~dBm, where the proposed scheme attains millimeter-level range accuracy, millimeter-per-second-level velocity accuracy, and angular errors below $0.01^\circ$, while the benchmarks incur larger errors in all four parameters. 
The gap to ESPRIT narrows at higher powers, but the proposed scheme shows no error floor for all scenarios. 
This is mainly attributed to the DD-based target separation and association prior to local refinement.

\begin{figure}[t]
 \centering
 \begin{minipage}[t]{0.49\columnwidth}
  \vspace{0pt}
  \centering
  \includegraphics[width=\linewidth]{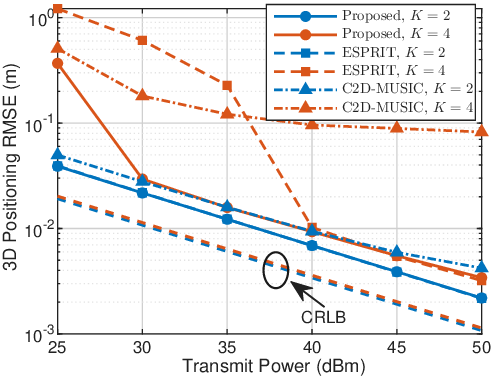}
  \caption{3D positioning RMSE versus transmit power with $R=4$.}
  \label{fig:position_rmse_vs_power}
 \end{minipage}\hfill
 \begin{minipage}[t]{0.49\columnwidth}
  \vspace{0pt}
  \centering
  \includegraphics[width=\linewidth]{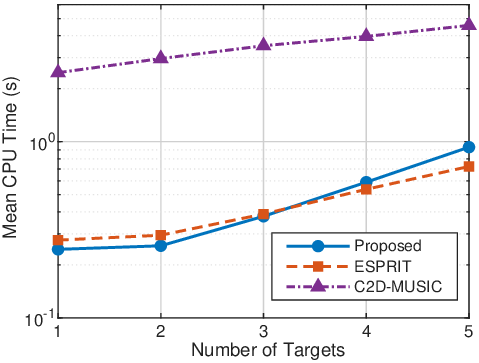}
  \caption{CPU time versus target number ($R=4$, $P_{\mathrm{t}}=35$~dBm).}
  \label{fig:cpu_time_vs_target_number}
 \end{minipage}
\end{figure}

Fig.~\ref{fig:position_rmse_vs_power} shows the 3D positioning RMSE obtained from the estimated ranges and angles. It is seen from the figure that for $K=2$, the proposed scheme achieves positioning accuracy comparable to ESPRIT. At 35~dBm, increasing $K$ from 2 to 4 raises the proposed RMSE only from about 1.2 to 1.6~cm, whereas the errors of ESPRIT and C2D-MUSIC increase more sharply. With four targets, the proposed positioning RMSE is approximately one order of magnitude lower than those of the two benchmarks.
With higher power, both ESPRIT and the proposed scheme attain comparable millimeter-level accuracy. 
The mid-power gap is consistent with the preservation of range and angle accuracy by the DD-based separation and association as the scene becomes more crowded.

\subsection{Computational Time}
\label{subsec:computational_time}

Fig.~\ref{fig:cpu_time_vs_target_number} reports the CPU time averaged over the fastest 95\% of runs.
The proposed algorithm has computational time comparable to ESPRIT and substantially lower than C2D-MUSIC across the considered target numbers.
By avoiding the dense 2D spectral searches of C2D-MUSIC, the proposed scheme combines the mid-power accuracy gains in Figs.~\ref{fig:parameter_rmse_vs_power} and \ref{fig:position_rmse_vs_power} with lower computational time.

\subsection{Impact of the Number of RF Chains}
\label{subsec:rf_chains}

\begin{figure}[t]
 \centering
 \includegraphics[width=0.98\columnwidth]{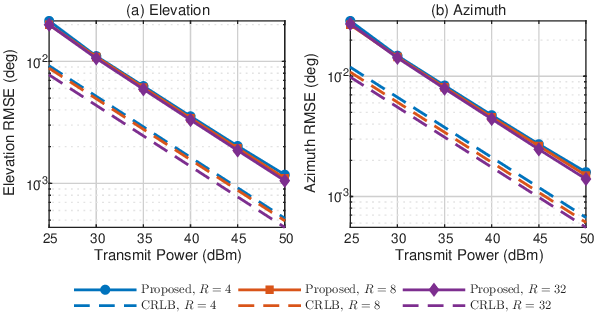}
 \caption{Angular RMSEs versus transmit power for $K=3$ and $R\in\{4,8,32\}$, with (a) elevation and (b) azimuth.}
 \label{fig:angle_vs_rf_chains}
\end{figure}

Fig.~\ref{fig:angle_vs_rf_chains} shows an overall reduction in both angular RMSEs from $R=4$ to $R=32$, although the improvement is modest relative to that obtained by increasing transmit power. The proposed curves follow the trend of their respective CRLBs, and even $R=4$ yields millidegree-level errors at high powers. 
Additional RF chains provide more spatial observations while retaining the same physical array aperture.
Therefore, the four-chain configuration already resolves the target angles accurately through the scanning and phase-center measurements.

\subsection{Ablation of Inter-Burst Doppler Refinement}
\label{subsec:wls_ablation}

\begin{figure}[t]
 \centering
 \includegraphics[width=0.58\columnwidth]{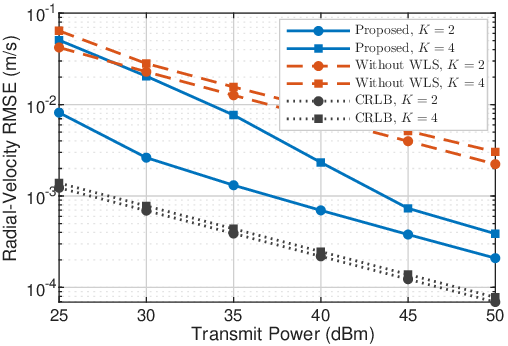}
 \caption{Radial-velocity RMSE versus transmit power with and without spatial-mask-aided WLS Doppler refinement ($R=4$, $K=2,4$).}
 \label{fig:ablation_wls}
\end{figure}

Fig.~\ref{fig:ablation_wls} compares the full estimator with a variant that retains the same global and local stages but omits the WLS refinement in \eqref{eq:fine_doppler}. At high powers, WLS reduces the radial-velocity RMSE by nearly an order of magnitude for both target numbers. 
This validates the effectiveness of the proposed Doppler refinement method.
The local $Q$-symbol estimate resolves Doppler ambiguity, whereas 
the longer temporal aperture provided by the target-specific inter-burst phase is effectively utilized to achieve near-CRLB estimation accuracy.

\section{Conclusion}
\label{sec:conclusion}

In this paper, we proposed a MIMO-OFDM ISAC framework that reuses downlink beam-scanning signals for multi-target sensing at BSs with few RF chains. A structured dual-tensor scheme was developed for global DD separation and local estimation of ranges, radial velocities, elevations, and azimuths. DD-based association linked the two stages, while spatial-mask-aided WLS refinement exploited inter-burst phases to improve Doppler accuracy. The joint multi-target CRLB was also derived to characterize estimation performance. Simulation results demonstrated centimeter- to millimeter-level 3D positioning with only four RF chains. The proposed scheme improved multi-target estimation accuracy over ESPRIT and C2D-MUSIC at moderate transmit powers, with runtime comparable to ESPRIT and substantially lower than C2D-MUSIC.

% External bibliography follows the format of our previous paper.
\bibliographystyle{IEEEtran}
\bibliography{IEEEabrv,ref}

\end{document}